\documentclass[a4paper,14pt,showpacs,preprintnumbers,amsmath,amssymb]{revtex4}

\usepackage{bm,hyperref}
\usepackage{amsmath}
\usepackage{lipsum}
\usepackage{graphicx,epstopdf}
\usepackage{amsfonts}
\usepackage{amssymb}
\usepackage{dcolumn}
\usepackage{amsmath}
\usepackage{soul}
\usepackage{color}

\usepackage{amssymb}
\usepackage{amsfonts}
\usepackage{amsmath}

\def\e{\begin{equation}}
\def\f{\end{equation}}
\def\=#1{\overline{\overline #1}}
\def\*#1{\overline{\overline{\overline #1}}}
\def\-#1{{\bf #1}}
\def\l#1{\label{eq:#1}}
\def\r#1{(\ref{eq:#1})}

\begin{document}
\title{Electromagnetic characterization of bianisotropic metasurfaces on refractive substrates: General theoretical framework}

\author{M. Albooyeh}
\email{corresponding author\\
mohammad.albooyeh@gmail.com}
\affiliation{%
 Department of Radio Science and Engineering, Aalto University, P.O. Box 13000, FI-00076 Aalto, Finland}%

\author{Sergei Tretyakov}
\affiliation{Department of Radio Science and Engineering, Aalto University, P.O. Box 13000, FI-00076 Aalto, Finland}

\author{Constantin Simovski$^{1,2}$}
\affiliation{$^1$Department of Radio Science and Engineering, Aalto University, P.O. Box 13000, FI-00076 Aalto, Finland}
\affiliation{$^2$Laboratory of Metamaterials, University for Information Technology, Mechanics and Optics (ITMO), 197101 St. Petersburg, Russia}

\begin{abstract}
We present a general methodology for electromagnetic homogenization and characterization of bianisotropic metasurfaces formed by regular or random arrangements of small arbitrary inclusions at the interface of two different isotropic media. These topologies can provide the most general electric and magnetic surface-bound dipolar responses to incident electromagnetic waves. The approach unites and generalizes the earlier theories developed independently by two research groups: the joint group of profs. Holloway and Kuester and the joint group of profs. Simovski and Tretyakov. We demonstrate the use of both formalisms in several example cases and discuss the differences between the two approaches. Furthermore, we generalize the known theories for the bianisotropic metasurfaces located at interfaces between two different media or on a substrate. The generalized framework allows characterization of both periodical or amorphous metasurfaces.  The developed analytical theory can be used in the analysis and synthesis of  metasurfaces for a wide variety of applications.

\bigskip
\noindent {Keywords:} electromagnetic characterization,
metamaterials, bianisotropy, plasmonic, characteristic
material parameters, metasurfaces

\bigskip
\noindent {Journal:} {JOPT}
\bigskip

\end{abstract}
\pacs{78.20.Ci, 42.70.Qs, 42.25.Gy, 73.20.Mf, 78.67.Bf}

 \maketitle
\section{Introduction}
\emph{Homogenization} means introducing some effective parameters which describe the response of a complicated system formed by a very large number of interacting elements. These parameters must model the material response as fully and accurately as is appropriate for the thought application. In electromagnetics, for volumetric (3D) materials and metamaterials, such parameters as permittivity and permeability are commonly used. Obviously, without this dramatic simplification of the problem, achieved by replacing considerations of microscopic distributions of currents and fields in each individual unit cell by studying homogenized distributions of currents and fields modeled by just a few material parameters, no realistic problem involving electromagnetic materials could be solved. For metasurfaces, permittivity and permeability loose their physical meaning, because they are defined in terms of volume-averaged fields and polarizations. It is clear that to model metasurfaces, one should average fields over the surface. Different effective parameters, such as sheet impedance or collective polarizability are used to model metasurfaces.

\emph{Characterization,} when applied in material science implies the use of external techniques to probe into the internal structure and properties of a material (see e.g. in Ref.~\cite{ref1_1}). In case of electromagnetic materials, characterization means determining the values of the effective parameters based on measured or simulated response of a material sample. Electromagnetic characterization of metamaterials (MMs) has been the focus of many studies, especially since double-negative composites had been introduced~\cite{ref1_1_1, ref1_1_2} in 2000-2001. In the current study, by electromagnetic characterization of any material including a MM, we mean finding some macroscopic parameters that can be used to predict the response of a material sample (e.g. a layer) to the electromagnetic waves. We can generally define a MM as a composite material with unusual electromagnetic properties offered by specific response of its constituents and their arrangement (see e.g. in Refs.~\cite{ref1_2, ref1_3, ref1_4, ref1_5}). Among the quickly growing MM literature, recently, the priority started to shift on studies of the 2D electrically/optically thin counterparts called metasurfaces (MSs), reviewed e.g. in Refs.~\cite{ref2_7, ref2_17}. MSs as well as bulk MMs, in the fields of external electromagnetic waves behave as effectively homogeneous structures, provided that the unit cells are sufficiently small. Therefore, it is reasonable to perform the characterization of an electrically/optically dense MS within the framework of a homogenization model; that is, a model describing the electromagnetic response of a MS in a condensed form. Indeed, we should be able to assign some macroscopic parameters to the MS in order to predict its behavior in response to an external electromagnetic wave.

These parameters, firstly, must be measurable experimentally or calculable numerically/analytically using available methods. Secondly, they should not depend on the incident wave polarization and direction of propagation (this is the homogeneity condition). Moreover, these parameters would depend on the topology of the constituents and their arrangement in the MS. Furthermore, they may be frequency dependent and this dependence is, as a rule, resonant. Of course, not all MSs can satisfy the second condition. It is clear that electrically/optically sparse planar arrays definitely do not belong to this category \cite{ref1_7}. As to electrically/optically dense arrays, those with resonant higher order multipolar responses rather than only dipolar responses are also not effectively homogeneous (at least we are not familiar with opposite examples). On the other hand, electrically/optically dense arrays with electric and magnetic dipole responses may be homogenized, being regular, aperiodic, random or even amorphous (see below).  Therefore, we concentrate on such -- dipolar -- MSs.

By taking the advantage of electromagnetic homogenization, we can predict the behavior of an array, consisting of many (theoretically of an infinite number of) inclusions, in response to electromagnetic fields independently of the complexity of each individual element and of their arrangement. This way, within a reasonable accuracy, we drastically reduce the calculation time and resources and gain clear physical understanding of complex electromagnetic phenomena. It is especially important on the stage of the optimization of a prospective MS.

To this end, in section \ref{chap2} we discuss the definitions and present a brief review of the electromagnetic characterization of MMs and MSs. We explain the failures of the traditional characterization approach when applied to MSs. We discuss two known approaches especially suggested for the characterization of MSs in 1990s-2000s and select one of them (though the alternative approach is also involved). We conclude that section by identifying the novel aspects  of the current study.

The third section is devoted to the theoretical mainstay of the present study. We start with the heuristic homogenization model of a metasurface located on a dielectric interface. After the rather special boundary conditions obtained for some MSs and separately for the transverse electric (TE) and transverse magnetic (TM) incident cases, we derive the general boundary conditions, valid for any polarization of the excitation field. Next, we introduce relations between the surface polarizations and the incident (one approach) or averaged (another approach) fields. In one case, we perform the characterization in terms of so-called collective polarizability dyadics. In another case, we deal with the surface susceptibility dyadics. Then, we present the most general algorithm to retrieve the characteristic parameters through two-dimensional reflection and transmission dyadics.

In section \ref{chap4}, we first present two explicit examples of theoretically investigated MSs in the previous section. The first one is a periodic array of plasmonic nano-spheres. We retrieve its effective susceptibilities and show their potential for predicting the reflection/transmission coefficients. The second one is an array of coupled plasmonic nano-patches positioned in a disordered fashion on a flat surface. We show that our retrieval approach for this random MS is as powerful as for a periodic MS. The restriction of our theory is a sufficiently small optical size of a unit cell (area per one particle). We continue this section with an example of a bianisotropic metasurface composed of split-ring resonators (SRRs). We finally compare two different approaches discussed in section \ref{chap2} for characterization of the proposed metasurface when we eventually conclude this study.

We hope that the techniques reported in this work, particularly the characterization of the bianisotropic response of metasurfaces with a substrate, may have a long standing impact in the literature, and mark an important milestone in the analytical theory of metasurfaces.

\section{A General Overview and History}\label{chap2}
\textbf{Metamaterial}: The most recent definition for the term ``metamaterial'' may be given~\cite{defmet} as following: A metamaterial is an electromagnetically homogeneous arrangement of artificial structural elements, designed to achieve advantageous and unusual electromagnetic properties.

\begin{figure}[h]
\centering
\includegraphics[width=0.49\textwidth,angle=0] {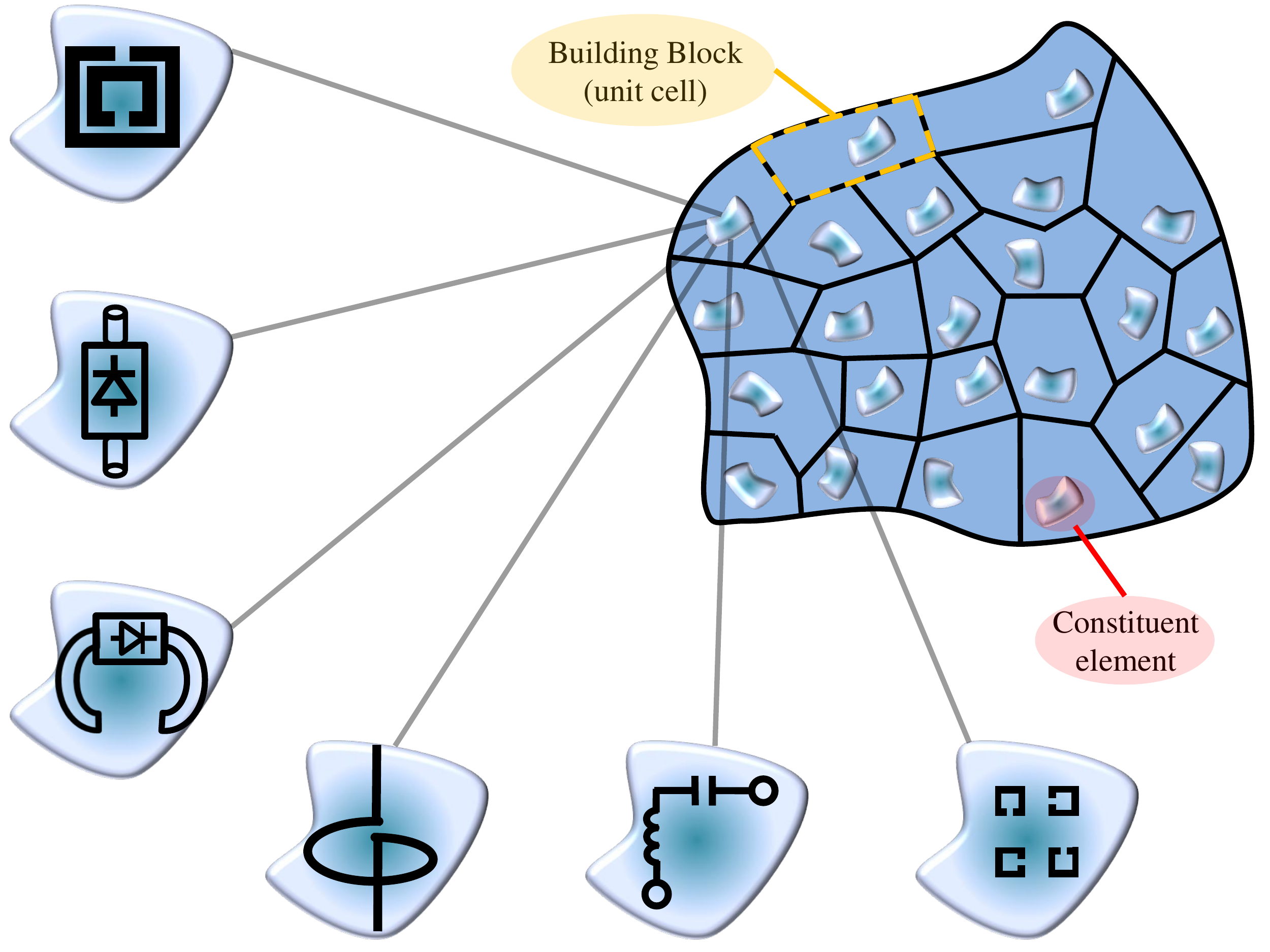}
\caption{The concept of metamaterial, building blocks, and constituent elements. The elements may generally be arranged amorphously. Also, the constituent elements may be any kind of electrical circuit or any other type of inclusions such as split-ring resonator, chiral particle, active element, etc. \cite{ref2_18, ref2_20}. The average size $a$ of a building block may be defined as the cubic root of its volume $V$; i.e., $a=\sqrt[3]{V}$.}
\label{FIG_CONC}
\end{figure}

Indeed, metamaterials are generally composed of smaller structural elements which are preferably assembled together. These elements play the same role as atoms and/or molecules do in ordinary, natural materials. However, the position and properties of each element may be desirably engineered to fulfill a specific goal not usually met in nature. The concept of a metamaterial prototype is illustrated in Fig.~\ref{FIG_CONC}.
\textbf{Characterization}: In electromagnetics, material characteristic parameters either refer to the constitutive (material) or to macroscopic wave parameters. Two most important macroscopic wave parameters are the wave impedance relating the amplitudes of the electric and magnetic fields in plane waves, and the refractive index relating the phase velocity of wave in the medium to the speed of light $c$. Material parameters are also macroscopic values and may be introduced only for effectively homogeneous media (materials). They relate the averaged electric and magnetic fields to the corresponding polarization densities. These parameters for bulk materials are usually two tensors: permittivity and permeability. In some materials called bianisotropic ones, there are two additional material tensors; that is, magnetoelectric and electromagnetic coupling tensors. For reciprocal bianisotropic media these two parameters reduce to one tensor which is usually split onto two simpler tensors; that is the so-called chirality and so-called omega-coupling tensors.
A comprehensive overview of bianisotropic media can be found in Ref.~\cite{ref5_0_2}.

The importance of the \textit{electromagnetic characterization} of materials, in general, and metamaterials, in particular, is that once we have found their characteristic parameters, we can predict the behavior of arbitrary-shaped samples in response to an arbitrary electromagnetic excitation. In particular, we can find how fast, with which power level, and in which direction an electromagnetic wave propagates in the presence of the target material in response to an incident wave.

\textbf{History}: Electromagnetic characterization of \textit{materials}; i.e., measuring/calculating their constitutive parameters dates back to the first half of twentieth century. Prof. Hippel \cite{ref2_11} in 1954, with the help of twenty-two contributors, has collected a plethora of different methods for measuring permittivity and permeability of various materials for a diverse frequency range. After that, in 1970s and with the advent of the computer and automatic test equipments, Nicolson, Ross, and Weir (NRW) have developed time-domain and frequency-domain techniques for the measurement of complex constitutive parameters~\cite{ref2_12, ref2_13}. These traditional methods, in the beginning of the twenty-first century, become the backbone of initial approach for characterization of \textit{metamaterials} in terms of their effective constitutive parameters by Smith \textit{et al.}~\cite{ref2_14}.

According to this approach, one uses the reflection and transmission data of a metamaterial slab, with a known thickness, in order to determine its effective permittivity and permeability as if it was a uniformly homogeneous bulk layer. Although this approach was appropriate and successful for ordinary \textit{materials}, it was subjected to ambiguities related to the correct definition of the front and rear surfaces of the sample for \textit{metamaterials}. Indeed, they resulted in non-physical frequency dispersion and wrong sign of the imaginary part of some retrieved material parameters (see e.g. in Ref.~\cite{ref1_5}). Moreover, this approach resulted in unique electromagnetic material parameters only for a sufficiently bulk metamaterial layer (five or more unit cells across its thickness). When applied for thin metamaterial slabs, the NRW approach fully failed; that is, the extracted material parameters strongly depended on the thickness of the test slab \cite{ref2_17, ref2_14, ref2_15, ref2_16}. While some methods have been used to resolve these ambiguities \cite{ref2_15}, none of them appeared to be general. Since the popularity of metasurfaces have grown due to their lower optical losses and easier manufacturing processes compared to their bulk counterparts, the lack of physically sound characterization techniques urged the specialists to search novel ways.

The joint team guided by Profs.~Holloway and Kuester was probably the first group who tried to solve the problem of metasurfaces' characterization from the theoretical standpoint. In work~\cite{ref2_17}, they shown the inapplicability of the NRW approach to metasurfaces.  First, they showed that the dependence of the retrieved material parameters on the physical sample thickness is conceptually related to cutoff modes excited at the junction of two different waveguides; i.e., the physical thickness of the metasurface is irrelevant for its characterization. They stated in Ref.~\cite{ref2_26} that ``The effective bulk material properties, which are determined from the same modified NRW approach used to analyze the bulk three-dimensional metamaterials, are not uniquely defined for metasurfaces. While the geometry of scatterers and the lattice constant (unit cell size) are uniquely defined, the thickness of the equivalent layer is obviously not.''

They then suggested an alternative characterization approach initially inspired by a classical work \cite{ref2_27} which was the first known paper treating the reflection and transmission of an incident plane wave for a crystalline or liquid half-space in terms of both bulk polarization of the medium and its surface polarization. In the review \cite{ref2_7}, the group of Holloway and Kuester presents many further attempts towards the response of polarizable surfaces to incident waves. Even more detailed overview on the history of this problem may be found in \cite{Albooyeh1, Albooyeh2}.

In works by Holloway-Kuester's team published during a decade (2003-2013) \cite{ref2_7, ref2_17}, a systematic analysis of  metasurfaces formed by resonant electrically and magnetically polarizable dipole scatterers resulted in a sound characterization technique. A magneto-dielectric metasurface was described via Generalized Sheet Transition Conditions (GSTCs) earlier introduced by Senior, Volakis\cite{ref2_27_1, ref2_27_2} and Idemen\cite{Idemen1, Idemen2, Idemen3, Idemen}. The presentation of metasurfaces using GSTCs are physical compared to an effective-bulk-medium model. Therefore, the metasurface after its homogenization in terms of surface polarizations (electric and magnetic) acts as an infinitesimally thin sheet of surface polarization currents which causes amplitude and phase changes in the macroscopic electric and magnetic fields. As a result, the electric and magnetic surface susceptibilities (or polarizabilities per unit area) that appear in the GSTCs are uniquely defined. These parameters can serve as characteristic metasurface parameters since they do not depend on the metasurface physical thickness. Moreover, these characteristic parameters turn out to be independent of the polarization of the external excitation \cite{ref2_7}.

Actually, even earlier than Holloway-Kuester's group started to develop their characterization technique, in works of another joint team, namely guided by Profs. Tretyakov and Simovski, an alternative model of metasurfaces had been developing since 1997 \cite{ref3_1_2, ref3_1_0, ref3_1_1, ref2_31, ref2_29, ref2_30, ref3_1_3}. This homogenization model was somewhat more general than Holloway-Kuester's one since it took into account the possible bianisotropy of scatterers. The approach was based on the concept of so-called collective polarizabilities of scattering particles. These polarizabilities relate the electric and magnetic polarizations of the unit cell with the \emph{incident} electric and magnetic fields at the array surface. These polarizabilities, like the \emph{surface susceptibilities} entering the GSTCs, could be retrieved from the reflection and transmission coefficients. Also, these polarizabilities were related with the surface impedances \cite{ref2_30, ref3_1_3}. Unfortunately, this model of metasurfaces was not sufficiently developed up to 2013; i.e., no feasible and robust retrieval algorithm was created based on it. Below, for brevity we refer this homogenization model as the ST model (developed in works of the group guided by Simovski and Tretyakov) and the model described in works of the group guided by Holloway and Kuester as the HK model.

In the following sections, we present the theory and results of electromagnetic characterization of general metasurfaces with linear response to electromagnetic fields. Basically, we generalize the HK model, though the ST model is also inspected, and we discuss the difference between these two approaches in what concerns the characterization of metasurfaces. Compared to the HK model, the present theory takes into account also:
\begin{enumerate}
 \item Possible bianisotropy of metasurfaces under characterization;
 \item Possible presence of the high-contrast substrate on which the scatterers are located;
 \item Two-side incidence of plane waves (important for bianisotropic metasurfaces when the reflected and/or transmitted fields are not necessarily the same for the illumination from different directions [see in Refs.~\onlinecite{ref3_1,Albooyeh1}]);
 \item Possible randomness of the metasurface.
\end{enumerate}
The novelty of our theory compared to the ST model results from:
\begin{enumerate}
 \item A rather simple and reliable algorithm of characterization;
 \item Possible presence of a high-contrast substrate;
 \item Possible randomness of the metasurface.
\end{enumerate}
In summary, in our theory, two significant drawbacks of both HK and ST models are overcome. That is, neglecting the influence of the substrate and the restriction to only periodical arrangements of scatterers. Moreover, we show that the ST and HK models represent two mathematically equivalent descriptions of metasurfaces.

We start the next section from formulating the general problem and adopted methodology. We then obtain the boundary conditions for a general metasurface. We continue with the representation of the reflected and transmitted fields in term of the incident (ST approach) or average fields (HK approach) using Maxwell's equations. We eventually establish a general algorithm to retrieve the effective parameters of the metasurface through the reflection and transmission coefficients.
\section{Theory}\label{chap3}

\subsection{Problem formulation}

\begin{figure}[h]
\centering
\includegraphics[width=0.49\textwidth,angle=0] {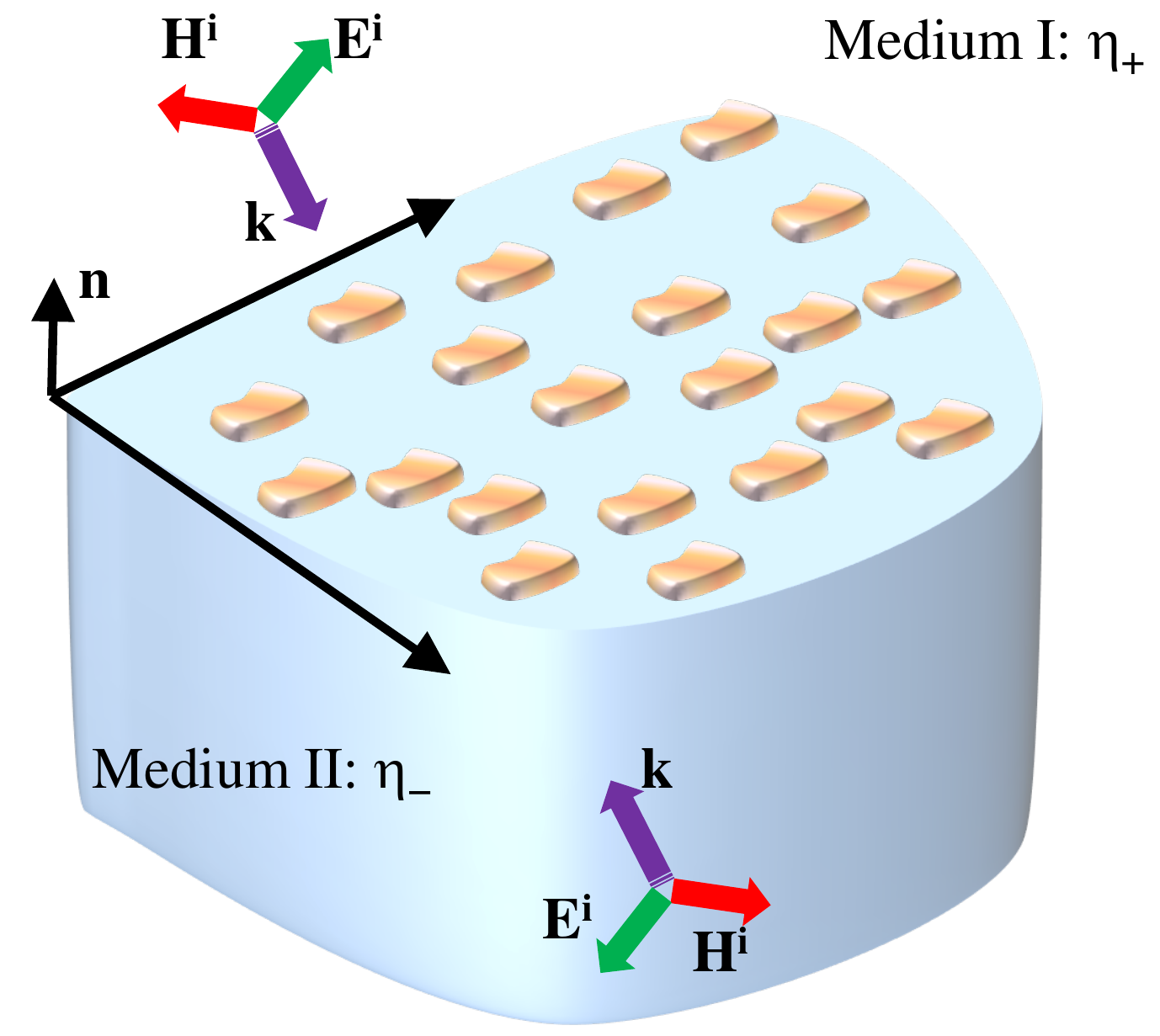}
\caption{Geometry of a metasurface composed of a dense planar array with an amorphous arrangement of its constituent elements located on top of a substrate with the characteristic impedances ``$\eta_-$''. The free-space characteristic impedance is denoted as $\eta_+$. Unit vector \textbf{n} is normal to the metasurface.}
\label{FIG_SCHtD}
\end{figure}

The problem is to characterize an electrically/optically dense planar array of resonant elements, with electric and possibly magnetic dipole responses to electromagnetic fields, using some macroscopic parameters. The array is supported by a semi-infinite dielectric substrate and after its homogenization is considered as a metasurface. The macroscopic parameters should be independent of the external excitation. They should only depend on the excitation frequency and physical parameters of the metasurface. These physical parameters e.g., the unit cell size and shape, the geometry of elements, and the materials of the elements and of the top (superstrate, $z>0$ in Fig.~\ref{FIG_SCHtD}) and bottom (substrate, $z<0$ in Fig.~\ref{FIG_SCHtD}) media, determine the frequency dispersion of the electromagnetic response.
This model should be unique; i.e., independent of the excitation, for an effectively homogeneous metasurface. The top medium for brevity is, in below, treated as free space, however, in our derivations we keep its wave impedance $\eta_+$ and refractive index $n_+$ in order to easily apply our results for two arbitrary isotropic media on two sides of the metasurface.

In our interpretation, metasurfaces can be electromagnetically homogenized via the averaging over each unit cell (Fig.~\ref{FIG_CONC}), under the assumption that the cells are sufficiently small compared to the operational wavelength. At the same time, the whole metasurface array is assumed to be extended enough, so that the characteristic parameters do not depend on the transversal sizes of the array.

Figure~\ref{FIG_SCHtD} shows a dense planar array of resonant particles which are randomly arranged in the general case. The arrangement may be periodic in special cases, however, as we will see below, our characterization model keeps valid both for amorphous (even fully random) and regular arrangements of constituents. In our derivations, we assume that the metasurface is excited by an electromagnetic plane wave which may impinge from either free space or the substrate. However, the resulting homogenized equations and characteristic parameters can be used for arbitrary excitation, provided that the incident field does not significantly change on the unit-cell size scale. With an external excitation by an electromagnetic wave, we may induce currents on each element of the metasurface. Each element, therefore, may re-radiate as a pair of electric and magnetic dipoles. Each dipole scatterer may interact with the other dipoles of the metasurface. According to \onlinecite{ref2_7, ref1_7}, we replace the array by a homogenous sheet with an electric and a magnetic surface polarizations ${\-{ P}}$ and ${\-{ M}}$. They represent the metasurface response to the external excitation. As a consequence, we model the metasurface response by surface currents which cause the field discontinuity\cite{Idemen} across the plane $z=0$.

In the next section, we derive the boundary conditions for such a metasurface. We demonstrate how the fields are related to each other, through the surface polarizations, on both sides of the surface $z=0$. The derivations are conducted for an arbitrary polarized electromagnetic wave.

\subsection{Boundary Conditions}

\begin{figure}[h]
\centering
\includegraphics[width=0.49\textwidth,angle=0] {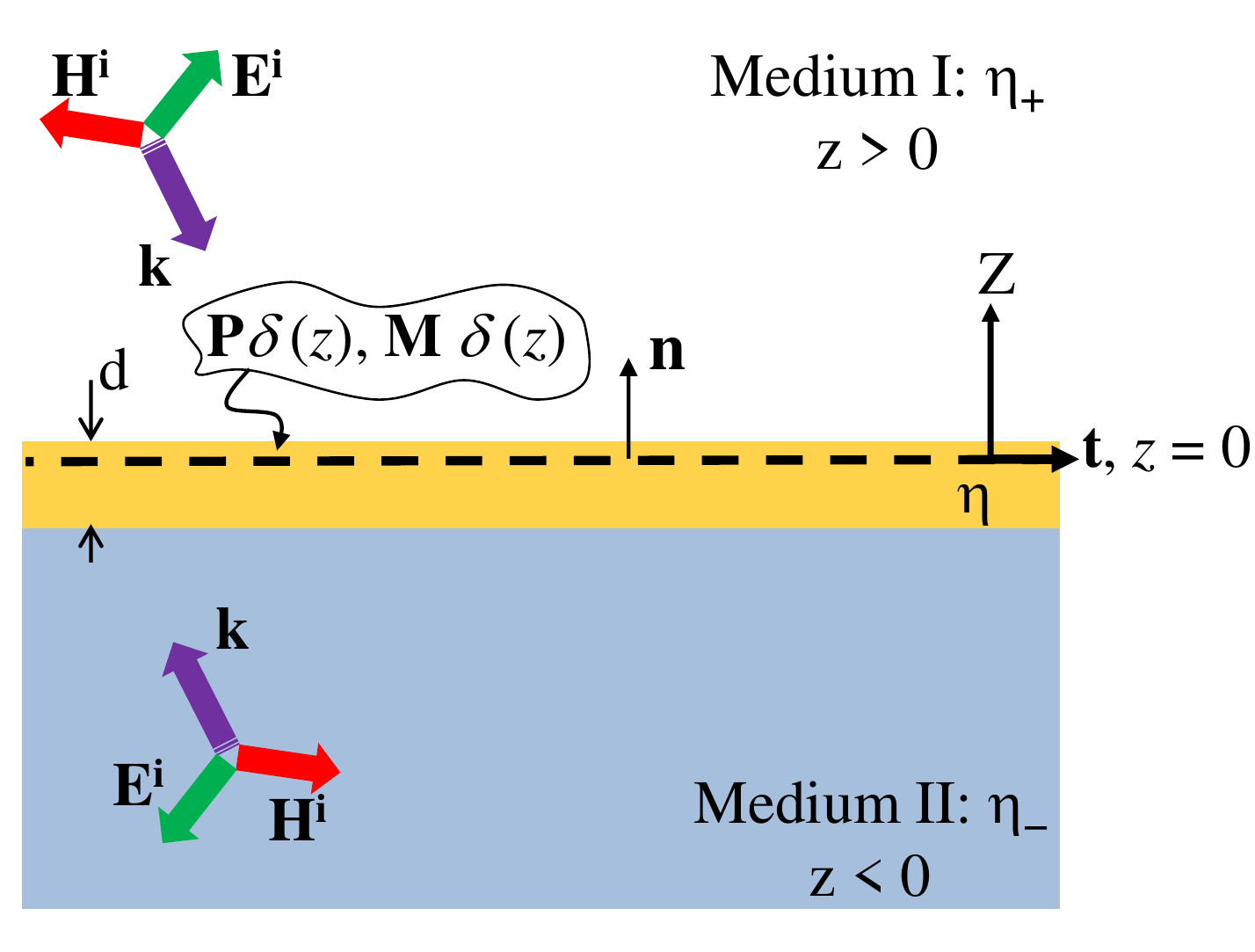}
\caption{Two-dimensional homogenized representation of a metasurface. The homogenization model is assumed to be a sheet maintaining  electric and magnetic bulk polarizations ``$\textbf{\textrm{P}}\delta(z)$'' and ``$\textbf{\textrm{M}}\delta(z)$'' which is located within a layer (not necessarily at the center of the layer) with the characteristic impedance $\eta$. The layer thickness is assumed to be at zero limit; i.e., $d \longrightarrow 0$. Moreover, it is assumed to be located at the interface of two different media with the characteristic impedances $\eta_+$ and $\eta_-$. The sheet (metasurface and the layer) may be illuminated from the forward ($z>0$) or backward ($z<0$) directions. Unit vector \textbf{n} is normal to the metasurface while unit vector \textbf{t} is tangent to the same surface.}
\label{FIG_SCH}
\end{figure}

Let us consider a metasurface located at an interface of two half-spaces with the characteristic impedances $\eta_+=\sqrt{\mu_+/\epsilon_+}$ (the upper one in Fig.~\ref{FIG_SCH}) and $\eta_-=\sqrt{\mu_-/\epsilon_-}$ (the lower one in Fig.~\ref{FIG_SCH}), where $\epsilon_{\pm}$ and $\mu_{\pm}$ are, respectively, the permittivity and permeability of the corresponding media. We assume that the constituent elements of the metasurface as well as the maximal unit cell size (see e.g. in Fig.~\ref{FIG_CONC}) are sufficiently small compared to the operational wavelength. Then the homogenization of the original metasurface makes sense. This homogenization implies averaging the electromagnetic fields as well as the microscopic electric and magnetic dipole polarizations in the plane $(x-y)$. For the regular metasurface with translational vector ${\bf a}=(a_x,a_y)$ (where $a_x$ and $a_y$ are array periods along axes $x$ and $y$, respectively), impinged by a plane wave with tangential wave vector ${\bf k}_\textrm{t}$ the tangential averaging can be defined as follows:
\e
{\bf A(r)}={e^{-j{\bf k}_\textrm{t}\cdot {\bf r}}\over S}\int\limits_{{\bf r}-{\bf a}/2}^{{\bf r}+{\bf a}/2}{\bf A}_{\rm micro}({\bf r'})e^{+j{\bf k}_\textrm{t}\cdot {\bf r'}}d^2{\bf r'}.
\l{def11}\f
Here ${\bf r}=(x,y)$ is the position vector in the array plane, $S=a_xa_y$ is the unit cell area, $\bf A$ is the averaged vector, and ${\bf A}_{\rm micro}$ is the microscopic vector. The time dependence in~\r{def11} is assumed to be $\exp {(j \omega t)}$. If the metasurface is illuminated by a finite source, e.g. a dipole antenna located in the far zone, the surface integration should have a more complex weight function than $\exp(+j{\bf k}_\textrm{t}\cdot {\bf r'})$. For random metasurfaces the surface integration in \r{def11} should be complemented by a statistical averaging. However, in  this paper we do not aim to consider these mathematical aspects. We simply imply that the tangential averaging of all vector fields allows us to get rid of microscopic oscillations in the plane $z=0$ keeping the dependence on $x$ and $y$ coordinates dictated by the incident field and the metasurface shape and size.

The surface averaging replaces the original metasurface by its homogenized model -- an infinitely thin sheet of electric and magnetic surface polarizations ${\-{ P}}\delta(z)$ and ${\-{ M}}\delta(z)$, respectively. Here $\delta(z)$ is the Dirac delta function, localized at the plane of the metasurface ($z=0$).
To derive the effective boundary conditions, we assume that the metasurface is located inside a thin layer of an isotropic magnetodielectric having  a small thickness $d$ which on the first stage is assumed to be finite. The material parameters of this host layer are denoted below $\epsilon$ and $\mu$. Sharing out this additional thin layer is relevant even if it is, in fact, a part of the substrate. It allows us to take into account the near-field coupling of the resonant array to the substrate. It is especially important for plasmonic metasurfaces whose elements create deeply subwavelength hot spots in the vicinity of these elements. It is a very important factor, which leads to such physical effects as substrate-induced binaisotropy \cite{Albooyeh1, Albooyeh2}. In the course of derivation, the thickness  $d$ is assumed to be negligibly small, since in real situations the metasurface response is determined by the inclusions. However, the polarization response of the inclusion hot spots depend on the properties of the support layer.

Following the procedure introduced in Ref.~\onlinecite{ref1_7} in  derivation of transition conditions for material layers, we start from Maxwell's equations for electric and magnetic fields and bulk electric and magnetic polarizations inside this additional layer:
\e
\-\nabla \times \-E = - j \omega \mu \left( \-H + {\-M \delta(z) \over {\mu}} \right)
\l{Efield}\f
\e
\-\nabla \times \-H =   j \omega \epsilon \left( \-E + {\-P\delta(z) \over {\epsilon}} \right)
\l{Hfield}\f
Here, $\-E$, $\-H$, $\-P$ and $\-M$ are, respectively, tangentially averaged electric and magnetic fields and surface electric and magnetic polarizations.

Next we decompose the fields and polarizations into the normal and transversal components in order to find the relations between the tangential electric and magnetic fields on two sides of the slab. We first define  the transversal (t) and normal (n) components of the fields ($\-E$, $\-H$) and polarizations ($\-P$, $\-M$):\begin{align}\l{TNDef}
  \nonumber \-E = \-E_\textrm{t}+{\-n}E_\textrm{n}, & \qquad \-H = \-H_\textrm{t}+{\-n}H_\textrm{n}, \\
  \-P = \-P_\textrm{t}+{\-n}P_\textrm{n}, & \qquad \-M = \-M_\textrm{t}+{\-n}M_\textrm{n}.
\end{align}
We also decomposed the nabla operator $\-\nabla$ \cite{ref1_7} into the respective parts:\e
\-\nabla = \-\nabla_\textrm{t} +{\partial \over {\partial z}}{\-n}.
\l{curlDef}\f
In \r{curlDef}, without loosing generality, we have assumed that the normal direction is along the $z$ axis. Now we may decompose \r{Efield} and \r{Hfield} using the definitions in \r{TNDef} and \r{curlDef}. After  simplifications, Maxwell's equations reduce into two sets:
\begin{eqnarray}\l{TComp}
  \nonumber \-\nabla_\textrm{t} \times {\-n}E_\textrm{n} + {\partial \over {\partial z}}({\-n} \times \-E_\textrm{t}) &=& - j \omega \mu \-H_\textrm{t} - j \omega \-M_\textrm{t}\delta(z), \\
  \-\nabla_\textrm{t} \times {\-n}H_\textrm{n} + {\partial \over {\partial z}}({\-n} \times \-H_\textrm{t})  &=& j \omega \epsilon \-E_\textrm{t} + j \omega \-P_\textrm{t} \delta(z),
\end{eqnarray}
with the transversal components of the polarizations and
\begin{eqnarray}\l{NComp}
  \nonumber {\-n}E_\textrm{n} &=& {1\over{j \omega \epsilon}} \-\nabla_\textrm{t} \times \-H_\textrm{t} - {\-n}{P_\textrm{n}\delta(z) \over {\epsilon}}, \\
  {\-n}H_\textrm{n} &=& -{1\over{j \omega \mu}} \-\nabla_\textrm{t} \times \-E_\textrm{t} - {\-n}{M_\textrm{n}\delta(z) \over {\mu}}\l{normal}
\end{eqnarray}
with their normal components. Next, substituting ${\-n}E_\textrm{n}$ and ${\-n}H_\textrm{n}$ from \r{NComp} into~\r{TComp} we obtain the following equations:
\begin{equation}\l{BT:E}
\begin{array}{l}
  \displaystyle
  {\partial \over {\partial z}}({\-n} \times \-E_\textrm{t}) = - j \omega \mu \-H_\textrm{t} - {1\over{j \omega \epsilon}} \-\nabla_\textrm{t} \times \-\nabla_\textrm{t} \times \-H_\textrm{t}
  \vspace*{.2cm}\\\displaystyle
  \hspace*{3.6cm}
  - j \omega \-M_\textrm{t}\delta(z)
  + \-\nabla_\textrm{t} \times {\-n}{P_\textrm{n} \delta(z) \over {\epsilon}},
\end{array}
\end{equation}
\begin{equation}\l{BT:H}
\begin{array}{l}
  \displaystyle
  {\partial \over {\partial z}}({\-n} \times \-H_\textrm{t}) = j \omega \epsilon \-E_\textrm{t} + {1\over{j \omega \mu}} \-\nabla_\textrm{t} \times \-\nabla_\textrm{t} \times \-E_\textrm{t}
  \vspace*{.2cm}\\\displaystyle
  \hspace*{3.6cm}
  + j \omega \-P_\textrm{t}\delta(z)
  + \-\nabla_\textrm{t} \times {\-n}{M_\textrm{n} \delta(z) \over {\mu}}.
\end{array}
\end{equation}
We next cross-multiply \r{BT:E} by $(-{\-n})$ to find the following set of equations:
\begin{equation}\l{BCT:E}
\begin{array}{l}
\displaystyle
{\partial \over {\partial z}} \-E_\textrm{t} = j \omega \mu \left[ \bar{\bar{I_\textrm{t}}}+{ \-\nabla_\textrm{t} \-\nabla_\textrm{t} \over {k^2}}\right] \cdot \left( {\-n} \times \-H_\textrm{t}\right)
\vspace*{.2cm}\\\displaystyle
\hspace*{2.2cm}
+ j \omega \left( {\-n} \times \-M_\textrm{t}\delta(z)\right) - \-\nabla_\textrm{t} {P_\textrm{n} \delta(z)\over {\epsilon}},
\end{array}
\end{equation}
\begin{equation}\l{BCT:H}
\begin{array}{l}
\displaystyle
{\partial \over {\partial z}} \left( {\-n} \times \-H_\textrm{t}\right) =  j \omega \epsilon \left[ \bar{\bar{I_\textrm{t}}}+{ \left( {\-n} \times \-\nabla_\textrm{t}\right)\left( {\-n} \times \-\nabla_\textrm{t}\right) \over {k^2}}\right] \cdot \-E_\textrm{t}
\vspace*{.2cm}\\\displaystyle
\hspace*{2.2cm}
+ j \omega \-P_\textrm{t} \delta(z) + \-\nabla_\textrm{t} \times {\-n} {M_\textrm{n} \delta(z)\over {\mu}}.
\end{array}
\end{equation}
Here $\bar{\bar{I_\textrm{t}}}$ is the two-dimensional unit dyadic; i.e., $\bar{\bar{I_\textrm{t}}}=\bar{\bar{I}}-{\-n}{\-n}$, and $k$ is the wave number inside the slab; i.e., $k = \omega \sqrt {\mu \epsilon}$. Also, $\-\nabla_\textrm{t} \-\nabla_\textrm{t}$ and $({\-n} \times \-\nabla_\textrm{t})({\-n} \times \-\nabla_\textrm{t})$ are two-dimensional dyadics with the following matrix representations:
$$
\-\nabla_\textrm{t} \-\nabla_\textrm{t} = \left( \begin{matrix} {\partial \over {\partial x}}\\ {\partial \over {\partial y}} \end{matrix} \right) \left( \begin{matrix} {\partial \over {\partial x}}&{\partial \over {\partial y}} \end{matrix} \right
) = \left( \begin{matrix} {\partial \over {\partial x}}{\partial \over {\partial x}} & {\partial \over {\partial x}}{\partial \over {\partial y}} \\ {\partial \over {\partial y}}{\partial \over {\partial x}} & {\partial \over {\partial y}}{\partial \over {\partial y}} \end{matrix} \right),
$$
and
$$
({\-n} \times \-\nabla_\textrm{t})({\-n} \times \-\nabla_\textrm{t}) = \left( \begin{matrix} -{\partial \over {\partial y}}\\ {\partial \over {\partial x}} \end{matrix} \right) \left( \begin{matrix} -{\partial \over {\partial y}}&{\partial \over {\partial x}} \end{matrix} \right) $$ $$ =\left( \begin{matrix} {\partial \over {\partial y}}{\partial \over {\partial y}} & -{\partial \over {\partial y}}{\partial \over {\partial x}} \\ -{\partial \over {\partial x}}{\partial \over {\partial y}} & {\partial \over {\partial x}}{\partial \over {\partial x}} \end{matrix} \right).
$$
Notice that for a plane wave solution, the spatial dependence is exponential; that is, $\exp{(-j\-k_\textrm{t} \cdot \-r)}$, and therefore we have $\-\nabla_\textrm{t} = -j \-k_\textrm{t}$ where $\-k_\textrm{t}$ is the tangential wave vector.

Next, we integrate equations~\r{BCT:E} and~\r{BCT:H} over the surface thickness and take the limit at $d\rightarrow 0$.
Integration of the derivatives in the left-hand side results in jumps of the corresponding vectors over the layer thickness. In the right-hand side, integrals of tangential field components tend to zero when  $d\rightarrow 0$, because the surface-averaged fields are everywhere finite, but the integration interval tends to zero. Thus, only the delta-function terms, which correspond to polarizations of metasurface inclusions located at $z=0$, survive in the right-hand side. As a result, we obtain the following vectorial forms of the generalized sheet transition conditions:
\begin{eqnarray}\l{STBC}
  \nonumber \-E^+_\textrm{t} - \-E^-_\textrm{t} &=& j \omega {\-n} \times {{\-{ M}}_\textrm{t}} - \-\nabla_\textrm{t} {{{ P}}_\textrm{n} \over {\epsilon}}, \\
  {\-n} \times \-H^+_\textrm{t} - {\-n} \times \-H^-_\textrm{t} &=& j \omega {{\-{ P}}_\textrm{t}} + \-\nabla_\textrm{t} \times {\-n} {{{ M}}_\textrm{n} \over {\mu}}.
\end{eqnarray}

Equations \r{STBC}, relating the jumps of transversal electric ($\-E^{\pm}_\textrm{t}$) and magnetic ($\-H^{\pm}_\textrm{t}$) fields across the sheet to the surface electric (${\-{P}}$) and magnetic (${\-{M}}$) polarizations, do not differ from the previously known GSTCs obtained in Ref. \onlinecite{ref2_7}. However, we have not only proven them for the case of the optical contrast $\eta_-\ne \eta_+$, but have also clarified the role of the substrate for the normal components of the polarization vectors. If the particles of the metasurface lie on top of the substrate, this component of the polarizations enters the GSTCs in the same form as in the conventional HK model. If the particles are submerged, this component is additionally divided by the relative permittivity (electric polarization) and permeability (magnetic one) of the substrate. If the particles are encapsulated by an electrically/optically thin layer of a third medium with material parameters $\epsilon$ and $\mu$, these components of the electric and magnetic polarizations are divisible by these parameters, respectively. As to the tangential polarizations, they enter the GSTCs in the same way for all three cases (particles are on top, submerged, or encapsulated), as in Ref. \onlinecite{ref2_7}. More importantly, in Section~\ref{effpol}, we show how the normal and tangential components of the surface polarizations are related to the tangential average fields.

At this step, we are ready to search for the reflected and transmitted fields in terms of these surface polarizations.

\subsection{Transverse components of the reflected and transmitted waves}

Let us now illuminate a  metasurface that is functionalized of being electrically and magnetically polarized with surface polarizations ${{\-{  P}}}$ and ${{\-{  M}}}$, by a plane electromagnetic wave ($\-\nabla_\textrm{t} = -j \-k_\textrm{t}$) and find the reflected and transmitted field assuming that we know the polarization vectors. We consider two cases: a) illumination from the forward direction which is denoted by the upper half space ($z > 0$) in Fig.~\ref{FIG_SCH} and b) illumination from the backward direction which corresponds to the lower half space ($z < 0$) in the same figure. The upper half space has the characteristic impedance $\eta_+$ (by default, free space) while for the lower half space we have $\eta_-$ (by default, a dielectric substrate). It is known (see e.g. Ref.~\onlinecite{ref1_7}) that for a plane wave we have:
\e
{\-n} \times \-H_\textrm{t} = {\mp}{\overline{\overline{Z}}}_{\pm}^{-1}\cdot{\-E_\textrm{t}},
\l{EH}\f
where the \scriptsize $\begin{array}{c}
                                 \textrm{top} \\
                                 \textrm{bottom}
                               \end{array}
$\normalsize sign before the right hand side denotes the waves traveling in the direction ${\pm}z$. In \r{EH}, $\overline{\overline{Z}}_{\pm}$ is the dyadic impedance of the corresponding media and denoted by \cite{ref1_7}:\e
\displaystyle
\overline{\overline{Z}}_{\pm} = Z^{TM}_{\pm}{ \-k_\textrm{t} \-k_\textrm{t} \over {k_{t}^2}}+ Z^{TE}_{\pm} { ({\-n} \times \-k_\textrm{t})({\-n} \times \-k_\textrm{t}) \over {k_{t}^2}},
\l{Zdyadic}\f
and\e
\displaystyle
Z^{TM}_{\pm} = \eta_{\pm} \sqrt{1-{{k_{t}^2}\over {{k_{\pm}^2}}}},\quad Z^{TE}_{\pm} = {\eta_{\pm} \over \sqrt{1-{{k_{t}^2}\over {{k_{\pm}^2}}}}},
\l{ZTMTE}\f
where we denote\e
\displaystyle
k_{n,\pm} = \sqrt{{k_{\pm}^2} - {k_{t}^2} },
\l{kdef}
\f
as the propagation constant of eigenwaves in the normal direction ${\-n}$, and $k_{t}$ is the tangential wave number, and $k_{\pm}= \omega \sqrt{\mu_{\pm}\epsilon_{\pm}}$. The tangential wave number is the same at all interfaces as dictated by the boundary conditions. It is defined in terms of the wavenumber of the medium from which the wave is coming ($\eta_+$ or $\eta_-$) and the incidence angle $\theta^\textrm{i}$ as $k_\textrm{t}=k_{\pm}\cos \theta^\textrm{i}$.

Let us now decompose the incident, reflected, and transmitted electric/magnetic fields into their transversal ($\-E^\textrm{i}_\textrm{t}$/$\-H^\textrm{i}_\textrm{t}$, $\-E^\textrm{r}_\textrm{t}$/$\-H^\textrm{r}_\textrm{t}$, $\-E^\textrm{t}_\textrm{t}$/$\-H^\textrm{t}_\textrm{t}$) and normal ($E^\textrm{i}_\textrm{n}$/$H^\textrm{i}_\textrm{n}$, $E^\textrm{r}_\textrm{n}$/$H^\textrm{r}_\textrm{n}$, $E^\textrm{t}_\textrm{n}$/$H^\textrm{t}_\textrm{n}$) components as following:\e
\-E^\textrm{i} = \-E^\textrm{i}_\textrm{t} + {\-n} E^\textrm{i}_\textrm{n}, \quad \-E^\textrm{r} = \-E^\textrm{r}_\textrm{t} + {\-n} E^\textrm{r}_\textrm{n}, \quad \-E^\textrm{t} = \-E^\textrm{t}_\textrm{t} + {\-n} E^\textrm{t}_\textrm{n},
\l{Einc}\f\e
\-H^\textrm{i} = \-H^\textrm{i}_\textrm{t} + {\-n} H^\textrm{i}_\textrm{n}, \quad \-H^\textrm{r} = \-H^\textrm{r}_\textrm{t} + {\-n} H^\textrm{r}_\textrm{n}, \quad \-H^\textrm{t} = \-H^\textrm{t}_\textrm{t} + {\-n} H^\textrm{t}_\textrm{n},
\l{Hinc}\f
Then, for the transversal components, in the case of illumination from the forward (top) direction, we have:\begin{eqnarray}
  \-E^+_\textrm{t} &=& \-E^\textrm{i}_\textrm{t} + \-E^\textrm{r}_\textrm{t} , \qquad \-E^-_\textrm{t} = \-E^\textrm{t}_\textrm{t}, \l{topEH:E} \\
  \-H^+_\textrm{t} &=& \-H^\textrm{i}_\textrm{t} + \-H^\textrm{r}_\textrm{t} , \qquad \-H^-_\textrm{t} = \-H^\textrm{t}_\textrm{t}, \l{topEH:H}
\end{eqnarray}
while for the illumination from the backward (bottom) direction, we have:\begin{eqnarray}
  \-E^+_\textrm{t} &=& \-E^\textrm{t}_\textrm{t} , \qquad \-E^-_\textrm{t} = \-E^\textrm{i}_\textrm{t} + \-E^\textrm{r}_\textrm{t}, \l{botEH:E} \\
  \-H^+_\textrm{t} &=& \-H^\textrm{t}_\textrm{t} , \qquad \-H^-_\textrm{t} = \-H^\textrm{i}_\textrm{t} + \-H^\textrm{r}_\textrm{t}. \l{botEH:H}
\end{eqnarray}
By using \r{EH} in \r{topEH:E}--\r{botEH:H} and then substituting $\-E^{\pm}_\textrm{t}$ and $\-H^{\pm}_\textrm{t}$ in the transversal boundary conditions \r{STBC}, one may find the following two sets of equations for the illumination from the forward direction:
\e
\-E^\textrm{i}_\textrm{t} + \-E^\textrm{r}_\textrm{t} - \-E^\textrm{t}_\textrm{t} =  j \omega {\-n} \times {{\-{  M}}_\textrm{t}} + j \-k_\textrm{t} {{{  P}}_\textrm{n} \over {\epsilon}} \l{topB:E},
\f
\e
{{\overline{\overline{Z}}_+^{-1}}} \cdot \-E^\textrm{i}_\textrm{t} - {{\overline{\overline{Z}}_+^{-1}}} \cdot \-E^\textrm{r}_\textrm{t} - {{\overline{\overline{Z}}_-^{-1}}} \cdot \-E^\textrm{t}_\textrm{t}=  j \omega {{\-{  P}}_\textrm{t}} - j \-k_\textrm{t} \times {\-n} {{{  M}}_\textrm{n} \over {\mu}}, \l{topB:H}
\f
and from the backward direction:
\e
\-E^\textrm{t}_\textrm{t} - \-E^\textrm{i}_\textrm{t} - \-E^\textrm{r}_\textrm{t} =  j \omega {\-n} \times {{\-{  M}}_\textrm{t}} + j \-k_\textrm{t} {{{  P}}_\textrm{n} \over {\epsilon}} \l{botB:E},
\f
\e
-{{\overline{\overline{Z}}_+^{-1}}} \cdot \-E^\textrm{t}_\textrm{t} + { {\overline{\overline{Z}}_-}} \cdot \-E^\textrm{i}_\textrm{t} - { {\overline{\overline{Z}}_-^{-1}}} \cdot \-E^\textrm{r}_\textrm{t}=  j \omega {{\-{  P}}_\textrm{t}} - j \-k_\textrm{t} \times {\-n} {{{  M}}_\textrm{n} \over {\mu}}, \l{botB:H}
\f
respectively.

The final step for finding the transversal components of the reflected $\-E^\textrm{r}_\textrm{t} $ and transmitted $\-E^\textrm{t}_\textrm{t}$ waves is to separately solve two sets of equations \r{topB:E}--\r{botB:H}. After some algebraic simplifications, the results for the transversal components of the reflected $\-E^\textrm{r}_\textrm{t}$ and transmitted $\-E^\textrm{t}_\textrm{t}$ electric fields read as:\begin{equation}
\begin{array}{l}
\displaystyle
\-E^\textrm{r}_\textrm{t} =- \left( \bar{\bar{I}}_\textrm{t} + {\overline{\overline{Z}}_{\mp}} {\overline{\overline{Z}}_{\pm}^{-1}} \right)^{-1} \left( \bar{\bar{I}}_\textrm{t} - {\overline{\overline{Z}}_{\mp}} {\overline{\overline{Z}}_{\pm}^{-1}} \right) \cdot \-E^\textrm{i}_\textrm{t}
\vspace*{.2cm}\\\displaystyle
\hspace*{0.9cm}
- \left( \bar{\bar{I}}_\textrm{t} + {\overline{\overline{Z}}_{\mp}} {\overline{\overline{Z}}_{\pm}^{-1}} \right)^{-1} \cdot \left[j \omega  \left( {\overline{\overline{Z}}_{\mp}} \cdot {{\-{  P}}_\textrm{t}} ~{\mp}~ {\-n} \times {{\-{  M}}_\textrm{t}} \right)
\right.\vspace*{.2cm}\\\displaystyle
\hspace*{0.9cm}\left.
{\mp} j \left( \-k_\textrm{t} {{{  P}}_\textrm{n} \over {\epsilon}} ~{\pm}~ {\overline{\overline{Z}}_{\pm}} \cdot \left({\-n} \times \-k_\textrm{t} \right) {{{  M}}_\textrm{n} \over {\mu}} \right) \right],
\end{array}\l{topbotBf:E}
\end{equation}
and
\begin{equation}
\begin{array}{l}
\displaystyle
\-E^\textrm{t}_\textrm{t} =2 \left( \bar{\bar{I}}_\textrm{t} + {\overline{\overline{Z}}_{\mp}} {\overline{\overline{Z}}_{\pm}^{-1}} \right)^{-1} \cdot \-E^\textrm{i}_\textrm{t}
\vspace*{.2cm}\\\displaystyle
\hspace*{0.9cm}
- \left( \bar{\bar{I}}_\textrm{t} + {\overline{\overline{Z}}_{\mp}} {\overline{\overline{Z}}_{\pm}^{-1}} \right)^{-1} \cdot \left[j \omega  \left( {\overline{\overline{Z}}_{\pm}} \cdot {{\-{  P}}_\textrm{t}} ~{\pm}~ {\-n} \times {{\-{  M}}_\textrm{t}} \right)
\right.\vspace*{.2cm}\\\displaystyle
\hspace*{0.9cm}\left.
{\pm} j \left( \-k_\textrm{t} {{{  P}}_\textrm{n} \over {\epsilon}} ~{\mp}~ {\overline{\overline{Z}}_{\mp}} \cdot \left( {\-n} \times \-k_\textrm{t} \right) {{{  M}}_\textrm{n} \over {\mu}} \right) \right],
\end{array}\l{topbotBf:H}
\end{equation}
where \scriptsize $\begin{array}{c}
                                 \textrm{top} \\
                                 \textrm{bottom}
                               \end{array}
$\normalsize sign in the brackets denotes for the illumination from \scriptsize $\begin{array}{c}
                                 \textrm{forward} \\
                                 \textrm{backward}
                               \end{array}
$\normalsize direction. In a special case, when the metasurface is located in a homogeneous host medium; i.e., $\overline{\overline{Z}}_+ = \overline{\overline{Z}}_- = \overline{\overline{Z}}$, we have:\begin{equation}
\begin{array}{l}
\displaystyle
\-E^\textrm{r}_\textrm{t} = - {1\over 2} \left[j \omega  \left( {\overline{\overline{Z}}} \cdot {{\-{  P}}_\textrm{t}} {\mp} {\-n} \times {{\-{  M}}_\textrm{t}} \right)
\right.\vspace*{.2cm}\\\displaystyle
\hspace*{0.9cm}\left.
{\mp} j \left( \-k_\textrm{t} {{{  P}}_\textrm{n} \over {\epsilon}} {\pm} {\overline{\overline{Z}}} \cdot \left({\-n} \times \-k_\textrm{t} \right) {{{  M}}_\textrm{n} \over {\mu}} \right) \right]
\end{array}\l{tbBf:E}
\end{equation}
and\begin{equation}
  \begin{array}{l}
  \displaystyle
     \-E^\textrm{t}_\textrm{t} = \-E^\textrm{i}_\textrm{t} - {1 \over 2} \left[j \omega  \left( {\overline{\overline{Z}}} \cdot {{\-{  P}}_\textrm{t}} ~{\pm}~ {\-n} \times {{\-{  M}}_\textrm{t}} \right)
     \right.\vspace*{.2cm}\\\displaystyle
\hspace*{0.9cm}\left.
     {\pm} j \left( \-k_\textrm{t} {{{  P}}_\textrm{n} \over {\epsilon}} ~{\mp}~ {\overline{\overline{Z}}} \cdot \left( {\-n} \times \-k_\textrm{t} \right) {{{  M}}_\textrm{n} \over {\mu}} \right) \right].
   \end{array}\l{tbBf:H}
\end{equation}
Notice that the equations~\r{topbotBf:E}--\r{tbBf:H} which present the reflected and transmitted fields in terms of the surface polarizations, are rather general. They may be applied for any special case of external illumination; for example, TE or TM-polarized  incident waves or a superposition of these two. More importantly, as we show later, all of these surface polarizations may be expressed in terms of the transversal electric field $\-E_\textrm{t}$ only.

In the next section, we will  present the relations between the surface polarizations ${\-{  P}}$ and ${\-{  M}}$ with the electromagnetic fields $\-E$ and $\-H$. Two different relations can be obtained: one corresponds to the HK approach and other one to the ST approach.

\subsection{Effective polarizability/susceptibility tensors: characteristic parameters of metasurfaces}\label{effpol}

The parameters which relate the fields to the surface polarizations are called polarizabilities or susceptibilities depending on the definition of the fields associated to the relations. Really, they are the constitutive parameters which do not change with different conditions of the external excitation and they only depend on the physical parameters of the metasurface; i.e., the shapes and sizes of the metasurface elements and unit cells. In the HK model, one uses the total fields whose different values on two sides of the metasurface are averaged.
In the ST model, one uses the incident field in the constitutive relations. Therefore, we may write the following constitutive relations between the fields and polarizations in the ST approach:
\begin{eqnarray}\l{constRP}
  \nonumber {\-{  P}} &=& \hat{\bar{\bar{\alpha}}}^{\textrm{ee}} \cdot \-E^\textrm{i}+\hat{\bar{\bar{\alpha}}}^{\textrm{em}} \cdot \-H^\textrm{i}, \\
  {\-{  M}} &=& \hat{\bar{\bar{\alpha}}}^{\textrm{me}} \cdot \-E^\textrm{i}+\hat{\bar{\bar{\alpha}}}^{\textrm{mm}} \cdot \-H^\textrm{i}, \l{our_collective}
\end{eqnarray}
where $\hat{\bar{\bar{\alpha}}}^{\textrm{ee}}$, $\hat{\bar{\bar{\alpha}}}^{\textrm{em}}$, $\hat{\bar{\bar{\alpha}}}^{\textrm{me}}$, and $\hat{\bar{\bar{\alpha}}}^{\textrm{mm}}$ are called \textit{collective polarizability} tensors: electric, magnetoelectric, electromagnetic, and magnetic ones, respectively. They, indeed, relate the incident fields to the same surface polarizations as we introduced above. We may alternatively define another set of characteristic parameters (the HK approach):
\begin{eqnarray}\l{constRS}
  \nonumber {\-{  P}} &=& \hat{\bar{\bar{\chi}}}^{\textrm{ee}} \cdot \-E^{\rm ave}+\hat{\bar{\bar{\chi}}}^{\textrm{em}} \cdot \-H^{\rm ave}, \\
  {\-{  M}} &=& \hat{\bar{\bar{\chi}}}^{\textrm{me}} \cdot \-E^{\rm ave}+\hat{\bar{\bar{\chi}}}^{\textrm{mm}} \cdot \-H^{\rm ave}. \l{HK_defin}
\end{eqnarray}
Here, $\hat{\bar{\bar{\chi}}}$ denotes \emph{surface susceptibility} tensors. In~\r{constRS}, all vectors in the form of $\-A^{\rm ave}$ refer to the \emph{average} fields, calculated as
\e
\-A^{\rm ave}= {{\-A^\textrm{i}+\-A^\textrm{r}+\-A^\textrm{t}}\over2},
\l{AveDef}\f
where superscripts $\textrm{i}$, $\textrm{r}$, and $\textrm{t}$ correspond to the incident, reflected, and transmitted wave fields, respectively (all taken at $z=0$). It is important not to confuse this averaging of the fields at the two sides of the surface with the surface averaging introduced above. All the fields in \r{AveDef} are already surface-averaged. Notice, that the incident, reflected, and transmitted fields satisfy Maxwell's equations in respective homogeneous media. Thus, the normal components of the incident and average fields can be expressed in terms of only transversal components of the same fields, using Eqs.~\r{normal} where the delta-function terms are absent. It means that although the metasurface inclusions can in general react to external fields polarized along all directions, it is enough to keep only tangential field components in  \r{our_collective} and \r{HK_defin}.

At this point, we have all necessary tools for calculation of the two-dimensional reflection and transmission dyadics $\bar{\bar{r}}$ and $\bar{\bar{t}}$ as functions of the effective polarizability/susceptibility tensors. We use the notations  $\-E^\textrm{r}_\textrm{t} = \bar{\bar{r}} \cdot \-E^\textrm{i}_\textrm{t}$ and $\-E^\textrm{t}_\textrm{t} = \bar{\bar{t}} \cdot \-E^\textrm{i}_\textrm{t}$. One may substitute the relations~\r{constRP} into equations~\r{topbotBf:E} and~\r{topbotBf:H} to find the reflection and transmission dyadics as functions of the effective polarizability tensors. Notice, when using~\r{constRS} as the constitutive relations, it is easier to substitute them into the boundary conditions~\r{topB:E}--\r{botB:H} and then solve these equations for $\bar{\bar{r}}$ and $\bar{\bar{t}}$, to find them as functions of the effective susceptibility tensors.

Before presenting the general approach and algorithm, we draw your attention to some important points related to calculations of the reflection and transmission dyadics from the obtained boundary conditions. In equations~\r{topbotBf:E} and~\r{topbotBf:H} we need to find ${{\-{  P}}_\textrm{t}}$ and ${{\-{  M}}_\textrm{t}}$ in terms of the incident/average field $\-E^{i}$/$\-E^{\rm ave}$. Let us first consider an arbitrary three-dimensional dyadic $\bar{\bar{A}}$ as following:\e
\bar{\bar{A}}=\left( \begin{matrix} A_{xx} & A_{xy} & A_{xz} \\ A_{yx} & A_{yy} & A_{yz} \\ A_{zx} & A_{zy} & A_{zz} \end{matrix} \right).
\l{DyaDef}\f
If we introduce the following dyadic, vectors, and scalar $\bar{\bar{A}}_\textrm{tt}$, $\bar{A}_\textrm{tn}$, $\bar{A}_\textrm{nt}$, and $A_\textrm{nn}$ formed by the components of the initial dyadic $\bar{\bar{A}}$ as following, $$
\bar{\bar{A}}_\textrm{tt}=\left( \begin{matrix} A_{xx} & A_{xy} \\ A_{yx} & A_{yy} \end{matrix} \right), \quad \bar{A}_\textrm{tn}=\left( \begin{matrix} A_{xz} \\ A_{yz} \end{matrix} \right),\l{DyaDef1}$$
\e \bar{A}_\textrm{nt}=\left( \begin{matrix} A_{zx} & A_{zy} \end{matrix} \right), \quad A_\textrm{nn}=A_{zz},
\l{DyaDef1}\f
we can present any three-dimensional dyadics $\bar{\bar{A}}$ as a combination of a two-dimensional dyadic $\bar{\bar{A}}_\textrm{tt}$, two two-component vectors $\bar{A}_\textrm{tn}$ and $\bar{A}_\textrm{nt}$, and a scalar value $A_\textrm{nn}$; that is,
\e
\bar{\bar{A}}=\left( \begin{matrix} \bar{\bar{A}}_\textrm{tt} & \bar{A}_\textrm{tn} \\ \bar{A}_\textrm{nt} & A_\textrm{nn} \end{matrix} \right).
\l{DyaDef3}\f
Now, using the definitions in~\r{DyaDef1} for the effective polarizability tensors and the definitions in~\r{TNDef}, we may write the following relations for the tangential and normal components of surface polarizations ${\-{  P}}$ and ${\-{  M}}$ from~\r{constRP} in terms of the transversal incident electric field $\-E_\textrm{t}^\textrm{i}$:\begin{widetext}
\begin{eqnarray}
  \nonumber {{\-{  P}}_\textrm{t}} &=& \hat{\bar{\bar{\alpha}}}^{\textrm{ee}}_\textrm{tt} \cdot \-E_{\textrm{t}}^\textrm{i} \mp {1\over {\omega \epsilon_{\pm} }} \hat{\bar{\alpha}}^{\textrm{ee}}_\textrm{tn} \-k_\textrm{t} \cdot \left( {\overline{\overline{Z}}_{\pm}^{-1}} \cdot \-E_\textrm{t}^\textrm{i} \right)
          \pm \hat{\bar{\bar{\alpha}}}^{\textrm{em}}_\textrm{tt} \cdot \left[ {\-n} \times \left( {\overline{\overline{Z}}_{\pm}^{-1}} \cdot \-E_\textrm{t}^\textrm{i} \right) \right]
          - {1\over {k_{n,\pm}}} \hat{\bar{\alpha}}^{\textrm{em}}_\textrm{tn} \-k_\textrm{t} \cdot \left[ {\-n} \times \left( {\overline{\overline{Z}}_{\pm}^{-1}} \cdot \-E_\textrm{t}^\textrm{i} \right) \right], \l{PMDyad:Pt} \\
  \nonumber \\
  \nonumber {{  P}}_\textrm{n} &=& \hat{\bar{\alpha}}^{\textrm{ee}}_\textrm{nt} \cdot \-E_{\textrm{t}}^\textrm{i} \mp {1\over {\omega \epsilon_{\pm} }} \hat{\alpha}^{\textrm{ee}}_\textrm{nn} \-k_\textrm{t} \cdot \left( {\overline{\overline{Z}}_{\pm}^{-1}} \cdot \-E_\textrm{t}^\textrm{i} \right)
         \pm \hat{\bar{\alpha}}^{\textrm{em}}_\textrm{nt} \cdot \left[ {\-n} \times \left( {\overline{\overline{Z}}_{\pm}^{-1}} \cdot \-E_\textrm{t}^\textrm{i} \right) \right]
          - {1\over {k_{n,\pm}}} \hat{\alpha}^{\textrm{em}}_\textrm{nn} \-k_\textrm{t} \cdot \left[ {\-n} \times \left( {\overline{\overline{Z}}_{\pm}^{-1}} \cdot \-E_\textrm{t}^\textrm{i} \right) \right], \l{PMDyad:Pn}\\
          \nonumber \\
          \nonumber {{\-{  M}}_\textrm{t}} &=& \hat{\bar{\bar{\alpha}}}^{\textrm{me}}_\textrm{tt} \cdot \-E_{\textrm{t}}^\textrm{i} \mp {1\over {\omega \epsilon_{\pm} }} \hat{\bar{\alpha}}^{\textrm{me}}_\textrm{tn} \-k_\textrm{t} \cdot \left( {\overline{\overline{Z}}_{\pm}^{-1}} \cdot \-E_\textrm{t}^\textrm{i} \right)
          \pm \hat{\bar{\bar{\alpha}}}^{\textrm{mm}}_\textrm{tt} \cdot \left[ {\-n} \times \left( {\overline{\overline{Z}}_{\pm}^{-1}} \cdot \-E_\textrm{t}^\textrm{i} \right) \right]
          - {1\over {k_{n,\pm}}} \hat{\bar{\alpha}}^{\textrm{mm}}_\textrm{tn} \-k_\textrm{t} \cdot \left[ {\-n} \times \left( {\overline{\overline{Z}}_{\pm}^{-1}} \cdot \-E_\textrm{t}^\textrm{i} \right) \right], \l{PMDyad:Mt} \\
          \nonumber \\
          {{  M}}_\textrm{n} &=& \hat{\bar{\alpha}}^{\textrm{me}}_\textrm{nt} \cdot \-E_{\textrm{t}}^\textrm{i} \mp {1\over {\omega \epsilon_{\pm} }} \hat{\alpha}^{\textrm{me}}_\textrm{nn} \-k_\textrm{t} \cdot \left( {\overline{\overline{Z}}_{\pm}^{-1}} \cdot \-E_\textrm{t}^\textrm{i} \right)
          \pm \hat{\bar{\alpha}}^{\textrm{mm}}_\textrm{nt} \cdot \left[ {\-n} \times \left( {\overline{\overline{Z}}_{\pm}^{-1}} \cdot \-E_\textrm{t}^\textrm{i} \right) \right]
          - {1\over {k_{n,\pm}}} \hat{\alpha}^{\textrm{mm}}_\textrm{nn} \-k_\textrm{t} \cdot \left[ {\-n} \times \left( {\overline{\overline{Z}}_{\pm}^{-1}} \cdot \-E_\textrm{t}^\textrm{i} \right) \right], \l{PMDyad:Mn}
\end{eqnarray}
\end{widetext}

If we use the HK approach instead of the ST approach, we can simply replace dyadics $\hat{\bar{\bar{\alpha}}}$ by $\hat{\bar{\bar{\chi}}}$, and $\-E_\textrm{t}^\textrm{i}$ by the tangential average field; i.e., $\-E_\textrm{t}^\textrm{i} \rightarrow (\-E_\textrm{t}^\textrm{i}+\-E_\textrm{t}^\textrm{r}+\-E_\textrm{t}^\textrm{t})/2$ in the above equations.

Here, following to \r{DyaDef1} all three-dimensional effective polarizability dyadics $\hat{\bar{\bar{\alpha}}}$ are defined as:\e
\hat{\bar{\bar{\alpha}}}=\left( \begin{matrix} \hat{\bar{\bar{\alpha}}}_\textrm{tt} & \hat{\bar{\alpha}}_\textrm{tn} \\ \hat{\bar{\alpha}}_\textrm{nt} & \hat{\alpha}_\textrm{nn} \end{matrix} \right).
\l{DyaDefalpha}\f
Again, in the above equations, the \scriptsize $\begin{array}{c}
                                 \textrm{top} \\
                                 \textrm{bottom}
                               \end{array}
$\normalsize sign in the brackets corresponds to the illumination from \scriptsize $\begin{array}{c}
                                 \textrm{forward} \\
                                 \textrm{backward}
                               \end{array}
$\normalsize direction at \scriptsize $\begin{array}{c}
                                 z>0 \\
                                 z<0
                               \end{array}
$\normalsize.

It is very important to note that we have presented all tangential as well as normal components of the surface polarizations $\-P$ and $\-M$ in terms of the transversal electric field component $\-E_\textrm{t}$. In order to obtain this representation in equations \r{PMDyad:Pt}, we have used Eqs.~\r{normal},~\r{EH}, and~\r{Zdyadic} with the substitution $\nabla_\textrm{t} =-j{\-k}_\textrm{t}$ and have removed the delta-function terms. Therefore, all components of the electromagnetic fields are expressed through the transversal electric field component $\-E_\textrm{t}$ in the plane $z=0$.; i.e.,
\e
E_\textrm{n} = \mp {1\over { \omega \epsilon_{\pm} }} {\-k_\textrm{t} \cdot \left( {\overline{\overline{Z}}_{\pm}^{-1}} \cdot \-E_\textrm{t} \right) }, \l{En_Et}
\f
\e
H_\textrm{n} = - {1\over {k_{n,\pm}}} \-k_\textrm{t} \cdot \left[ {\-n} \times \left( {\overline{\overline{Z}}_{\pm}^{-1}} \cdot \-E_\textrm{t} \right) \right], \l{Hn_Et}
\f
\e
\-H_\textrm{t} = \pm {\-n} \times \left( {\overline{\overline{Z}}_{\pm}^{-1}} \cdot \-E_\textrm{t} \right). \l{Ht_Et}
\f
Here ${\overline{\overline{Z}}_{\pm}}$ and $k_{n,\pm}$ are defined in \r{Zdyadic} and \r{kdef}, respectively. This formalism makes the calculation of the reflected and transmitted electric fields and hence the two-dimensional reflection and transmission dyadics $\bar{\bar{r}}$ and $\bar{\bar{t}}$ general and compact.

The final step is to find the reflection and transmission dyadics relating $\-E_\textrm{t}^\textrm{r}$ and $\-E_\textrm{t}^\textrm{t}$ to $\-E_\textrm{t}^\textrm{i}$. This can be performed by substitution of the surface polarizations found in~\r{PMDyad:Pt} into the formulas for the reflected~\r{topbotBf:E} and transmitted~\r{topbotBf:H} electric fields. We omit the details of this procedure since one may find them in Refs.~\onlinecite{Albooyeh1, Albooyeh2, Albooyeh4} for different special cases.

We next present the general procedure of parameter retrieval suitable for both sets of characteristic parameters: collective polarizability and effective susceptibility tensors.

\subsection{General methodology}\label{Meth}

So far, we have obtained the relation between transversal components of the reflected $\-E_\textrm{t}^\textrm{r}$ and transmitted $\-E_\textrm{t}^\textrm{t}$ fields and the incident/average field $\-E_\textrm{t}^\textrm{i}$/$\-E_\textrm{t}^{\rm ave}$. In this section, we present the general algorithm of metasurface characterization. As we have already discussed, we emphasize that the presented formulas are suitable for the general case of an arbitrary polarized incident wave and arbitrary optical contrast between the medium of incidence and medium of transmission. Moreover, the metasurface may be bianisotropic with arbitrary anisotropy of electric, magnetic, and magneto-electric coupling tensors. Furthermore, the metasurface may not only possess tangential surface polarization components but also may include normal polarizations as well. Some special cases are studied in Refs.~\onlinecite{Albooyeh1, Albooyeh2, Albooyeh4, Albooyeh3, Albooyeh5, Albooyeh6}. For simplicity, we adopt below the notations of the ST approach. However, one may also choose the HK notations as we did for special cases in Refs.~\onlinecite{Albooyeh1, Albooyeh2}. The step-by-step algorithm to characterize the metasurfaces, therefore, is as following:\begin{enumerate}
\item First, we assume that the metasurface may be characterized by collective polarizabilities (electric: $\hat{\bar{\bar{\alpha}}}^{\textrm{ee}}$, magnetic: $\hat{\bar{\bar{\alpha}}}^{\textrm{mm}}$, magnetoelectric: $\hat{\bar{\bar{\alpha}}}^{\textrm{em}}$, and electromagnetic: $\hat{\bar{\bar{\alpha}}}^{\textrm{me}}$). These parameters relate the surface polarizations ${{\-{  P}}}$ and ${{\-{  M}}}$ to the incident transversal electric field as in \r{PMDyad:Mn}. We may briefly write these equations in the following form:
    \e
    {{\-{  P}}}, {{\-{  M}}} = f_1,f_2(\hat{\bar{\bar{\alpha}}}, \textbf{\textrm{k}}_\textrm{t}, \textbf{\textrm{E}}_\textrm{t}^\textrm{i}).
    \l{d1}\f
Here $f_1(\cdot)$ and $f_2(\cdot)$ define functions.
\item Then, we use the generalized sheet boundary conditions \r{STBC} and relate the reflected and transmitted fields to the incident field and the surface polarizations ${{\-{  P}}}$ and ${{\-{  M}}}$ as in~\r{topbotBf:E} and~\r{topbotBf:H}. In a simple form, these relations may be presented as:
    \e
    \-E^\textrm{r}_\textrm{t}, \-E^\textrm{t}_\textrm{t}= g_1,g_2(\textbf{\textrm{E}}^\textrm{i}_\textrm{t}, {{\-{  P}}}, {{\-{  M}}}).
    \l{d2}\f
Here $g_1(\cdot)$ $g_2(\cdot)$ denote two different functions.
\item Next, combining the first two items, one finds the reflection and transmission dyadics $\bar{\bar{r}}$ and $\bar{\bar{t}}$ as functions of the polarizability tensors and the vectorial operator $\-k_\textrm{t}$; i.e.;
    \e\-E^\textrm{r}_\textrm{t}, \-E^\textrm{t}_\textrm{t} = \bar{\bar{r}}, \bar{\bar{t}}(\hat{\bar{\bar{\alpha}}}, \textbf{\textrm{k}}_\textrm{t}) \cdot \textbf{\textrm{E}}^\textrm{i}_\textrm{t}.
    \l{d3}\f
Notice, as we have mentioned earlier, in the plane-wave representation $\-\nabla_\textrm{t} = -j\-k_\textrm{t}$. This implies that for the normal illumination $\-k_\textrm{t}=0$, while for oblique incidence, reflection and transmission dyadics are functions of the incident angle $\theta^\textrm{i}$: $k_\textrm{t}=k_{\pm}\cos\theta^\textrm{i}$.

 \item To characterize the metasurface based on reflection-transmission measurements or simulations, we express the characteristic parameters $\hat{\bar{\bar{\alpha}}}$ through the reflection and transmission dyadics, inverting equation~\r{d3}. It is not so simple because each polarizability tensor is a three-dimensional matrix which has generally $9$ components, and we generally need to determine $36$ complex scalars. However, we may have only $4$ equations for each of the two-dimensional tensors $\bar{\bar{r}}$ and $\bar{\bar{t}}$. Therefore, with one incidence case we may totally have only $8$ equations for $36$ unknowns. Nevertheless, we may do the same experiment for different incident angles, say for 5 angles, and obtain 40 numerical equations to reliably solve for all $36$ polarizability components which do not depend on the incidence angle. Yet, in many practical cases, depending on the physical shape of the metasurface elements and the excitation polarization, many of these scalar polarizability components are either negligibly small or can be related to one another by the reciprocity conditions or due to the symmetry of the structure. Therefore, we rarely need to solve for all 36 unknowns. For most important functional metasurfaces used for optical applications (e.g. see in Refs.~\onlinecite{Albooyeh1, Albooyeh2, Albooyeh4, Albooyeh3, Albooyeh5, Albooyeh6}) we only need to solve for $3$ or $4$ components from 36. Therefore, we only need to know the reflection/transmission coefficients for $2$ or $3$ incident angles. Finally, the solution to the characteristic parameters $\hat{\bar{\bar{\alpha}}}$ reads as:
    \e
    \hat{\bar{\bar{\alpha}}}= h(\bar{\bar{r}}(\theta), \bar{\bar{t}}(\theta), \textbf{\textrm{k}}_\textrm{t}).
    \l{d4}\f
    Here $h(\cdot)$ represents a function. Two-dimensional tensors $\bar{\bar{r}}(\theta)$ and $\bar{\bar{t}}(\theta)$ stand for the transversal reflection and transmission dyadics for an arbitrary incident angle $\theta$.

\item The final step of the algorithm is to validate the solution. One may insert the retrieved polarizabilities from \r{d4} into \r{d3} in order to find the reflection/transmission coefficients for other incident angles [different from those which were used for the parameter retrieval in~\r{d4}] and compare the result with reflection and transmission coefficients numerically simulated or experimentally measured for these test angles.
\end{enumerate}

Before going to the next section, it is worth to note that we have also developed a circuit model for substrated metasurfaces in Ref.~\onlinecite{Albooyeh5}. The drawback of the model in Ref.~\onlinecite{Albooyeh5} is its suitability for only one-direction incidence (different circuits for the illumination in the forward and in the backward directions). The equivalent circuit of Ref.~\onlinecite{Albooyeh5} is $\Gamma$-shaped; i.e., comprises only two equivalent lumped impedances, whereas the general circuit model applicable for both illumination cases is either the $T$-shaped or $\Pi$-shaped schemes comprising three lumped impedances \cite{Sergei, ref5_0_5}. The general idea is to use the concept of voltages and currents instead of the transversal electric and magnetic fields. Besides, effective polarizabilities/susceptibilities must be replaced by the effective sheet impedances and/or admittances. This may simplify the study especially when we deal with finite-thickness substrates or with piece-wise homogeneous (step inhomogeneity) metasurfaces.

In the next section, we consider some examples and show the applicability of the proposed characterization technique. The examples correspond to optical frequencies where the need in this technique is especially keen. We only present final results and present some physical discussions about them.

\let\cleardoublepage\clearpage
\section{Practical Examples}\label{chap4}
In the previous section, we presented a general approach to the characterization of metasurfaces. However, solving equation~\r{d4} to find the metasurface characteristic parameters (collective polarizabilities or surface susceptibilities) in the general case, is very complicated and rarely needed. It is reasonable to  reduce the problem complexities and simplify its solution. In the upcoming sections, we present different examples and show how a priori knowledge of the problem may simplify the study. In the first two examples we have adopted the HK model while in the last example we present a comparison between HK and ST approaches.

\subsection{A planar array of plasmonic nano-spheres: a periodic metasurface}\label{periMS}

\begin{figure}[h]
\centering
\includegraphics[width=0.49\textwidth,angle=0] {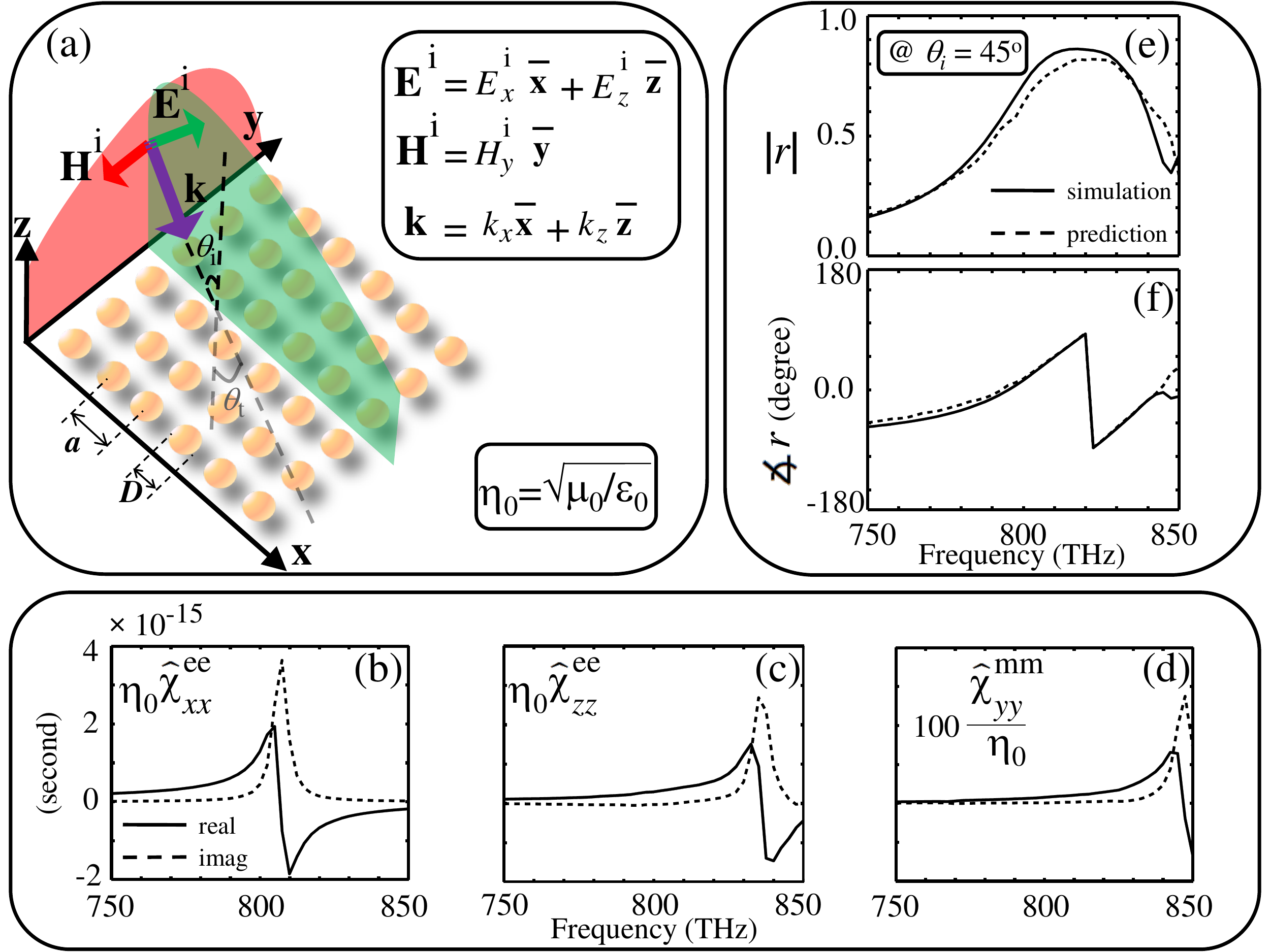}
\caption{(a) A planar periodic array of silver plasmonic nano-spheres positioned at the \textbf{xy}--plane in free space, where $\epsilon_0$ and $\mu_0$ are the permittivity and permeability of the free space, respectively. The array is excited by a TM-polarized electromagnetic plane wave. The diameter of each sphere is ${\textit{D}}=40$~nm while the cell size is ${\textit{a}}=80$~nm. (b) Tangential electric (c) Normal electric and (d) Tangential magnetic effective susceptibility components of the metasurface retrieved from the reflection and transmission data at $\theta_i=0$ and $10^\circ$. (e) Predicted and simulated results for the amplitude of the reflection coefficient of the proposed metasurface for an incidence of $\theta_i=45^\circ$. (f) The same plot as in (e) for the phase of the reflection coefficient. The material data for silver is taken from Ref.~\onlinecite{ref4_1}. Notice that the time dependence is assumed to be $\exp{(-i\omega t)}$ in these calculations.}
\label{FIG_Plas_Sph}
\end{figure}

Let us consider a dense planar periodic array of \textit{non-bianisotropic} plasmonic nano-spheres located in free space at $z=0$ as in Fig.~\ref{FIG_Plas_Sph}(a).
This array may be considered as a metasurface since the particles and the period of the array are enough smaller than the operational wavelength; that is, ${\textit{D}}=0.1 \lambda_0$ and ${\textit{a}}=0.2 \lambda_0$, where $\lambda_0$ is the central operational wavelength in the host medium (here, free space). Moreover, the response of this composite sheet is in the frequency band of its inclusions' resonances which implies a metasurface response.

According to the geometry of the problem, the general susceptibility tensors may be written as:\e
\hat{\bar{\bar{\chi}}}^{\textrm{ee}}= \left( \begin{matrix} \hat{\bar{\bar{\chi}}}_\textrm{tt}^{\textrm{ee}} & \hat{\bar{\chi}}_\textrm{tn}^{\textrm{ee}} \\ \hat{\bar{\chi}}_\textrm{nt}^{\textrm{ee}} & \hat{\chi}_\textrm{nn}^{\textrm{ee}} \end{matrix} \right) = \left( \begin{matrix} \hat{\chi}_{xx}^{\textrm{ee}} & \hat{\chi}_{xy}^{\textrm{ee}} & \hat{\chi}_{xz}^{\textrm{ee}} \\ \hat{\chi}_{yx}^{\textrm{ee}} & \hat{\chi}_{yy}^{\textrm{ee}} & \hat{\chi}_{yz}^{\textrm{ee}} \\ \hat{\chi}_{zx}^{\textrm{ee}} & \hat{\chi}_{zy}^{\textrm{ee}} & \hat{\chi}_{zz}^{\textrm{ee}} \end{matrix} \right),
\l{DefTensEE}\f
\e
\hat{\bar{\bar{\chi}}}^{\textrm{em}}= \left( \begin{matrix} \hat{\bar{\bar{\chi}}}_\textrm{tt}^{\textrm{em}} & \hat{\bar{\chi}}_\textrm{tn}^{\textrm{em}} \\ \hat{\bar{\chi}}_\textrm{nt}^{\textrm{em}} & \hat{\chi}_\textrm{nn}^{\textrm{em}} \end{matrix} \right)=\left( \begin{matrix} \hat{\chi}_{xx}^{\textrm{em}} & \hat{\chi}_{xy}^{\textrm{em}} & \hat{\chi}_{xz}^{\textrm{em}} \\ \hat{\chi}_{yx}^{\textrm{em}} & \hat{\chi}_{yy}^{\textrm{em}} & \hat{\chi}_{yz}^{\textrm{em}} \\ \hat{\chi}_{zx}^{\textrm{em}} & \hat{\chi}_{zy}^{\textrm{em}} & \hat{\chi}_{zz}^{\textrm{em}} \end{matrix} \right),
\l{DefTensEM}\f
\e
\hat{\bar{\bar{\chi}}}^{\textrm{me}}= \left( \begin{matrix} \hat{\bar{\bar{\chi}}}_\textrm{tt}^{\textrm{me}} & \hat{\bar{\chi}}_\textrm{tn}^{\textrm{me}} \\ \hat{\bar{\chi}}_\textrm{nt}^{\textrm{me}} & \hat{\chi}_\textrm{nn}^{\textrm{me}} \end{matrix} \right)=\left( \begin{matrix} \hat{\chi}_{xx}^{\textrm{me}} & \hat{\chi}_{xy}^{\textrm{me}} & \hat{\chi}_{xz}^{\textrm{me}} \\ \hat{\chi}_{yx}^{\textrm{me}} & \hat{\chi}_{yy}^{\textrm{me}} & \hat{\chi}_{yz}^{\textrm{me}} \\ \hat{\chi}_{zx}^{\textrm{me}} & \hat{\chi}_{zy}^{\textrm{me}} & \hat{\chi}_{zz}^{\textrm{me}} \end{matrix} \right),
\l{DefTensME}\f
and
\e
\hat{\bar{\bar{\chi}}}^{\textrm{mm}}= \left( \begin{matrix} \hat{\bar{\bar{\chi}}}_\textrm{tt}^{\textrm{mm}} & \hat{\bar{\chi}}_\textrm{tn}^{\textrm{mm}} \\ \hat{\bar{\chi}}_\textrm{nt}^{\textrm{mm}} & \hat{\chi}_\textrm{nn}^{\textrm{mm}} \end{matrix} \right)=\left( \begin{matrix} \hat{\chi}_{xx}^{\textrm{mm}} & \hat{\chi}_{xy}^{\textrm{mm}} & \hat{\chi}_{xz}^{\textrm{mm}} \\ \hat{\chi}_{yx}^{\textrm{mm}} & \hat{\chi}_{yy}^{\textrm{mm}} & \hat{\chi}_{yz}^{\textrm{mm}} \\ \hat{\chi}_{zx}^{\textrm{mm}} & \hat{\chi}_{zy}^{\textrm{mm}} & \hat{\chi}_{zz}^{\textrm{mm}} \end{matrix} \right).
\l{DefTensMM}\f However, as we mentioned earlier, there are some physical considerations which can help to reduce the problem complexity. At this point, we discuss these issues based on our prior knowledge about the proposed problem:\begin{enumerate}
  \item We are going to analyze a metasurface in the resonance band of its constituent elements; i.e., spheres. The particles are assumed to be made of silver which implies strong electric resonance. Therefore, the magnetic resonant response is negligible compared to the electric one; that is $\hat{\bar{\bar{\chi}}}^{\textrm{mm}} \approx 0$. However, we do not neglect the magnetic response in order to prove this issue. Nevertheless, neglecting this magnetic resonance causes a much easier solution.
  \item There is no physical reason for bianisotropy in the structure. Therefore, both bianisotropic tensors $\hat{\bar{\bar{\chi}}}^{\textrm{em}}$ and $\hat{\bar{\bar{\chi}}}^{\textrm{me}}$ must vanish.
  \item The particles are reciprocal and no polarization transformation can take place. This means that all cross components of the electric and magnetic susceptibility tensors $\hat{\bar{\bar{\chi}}}^{\textrm{ee}}$ and $\hat{\bar{\bar{\chi}}}^{\textrm{mm}}$ are zero; i.e., $\hat{\bar{\chi}}_\textrm{tn}^{\textrm{ee}} =$ $ \hat{\bar{\chi}}_\textrm{tn}^{\textrm{mm}} =$ $ \hat{\bar{\chi}}_\textrm{nt}^{\textrm{ee}} =$ $ \hat{\bar{\chi}}_\textrm{nt}^{\textrm{mm}} =$ $ \hat{\chi}_{xy}^{\textrm{ee}} =$ $ \hat{\chi}_{xy}^{\textrm{mm}} =$ $ \hat{\chi}_{yx}^{\textrm{ee}} =$ $ \hat{\chi}_{yx}^{\textrm{mm}} = 0$. As a result, the cross components of reflection and transmission tensors are also zero; that is, $r_{xy}=$ $r_{yx}=$ $t_{xy}=$ $t_{yx}=0$.
  \item The geometry of the structure and its excitation imposes some other conditions on the susceptibilities. With the TM-polarization of the incident field, there is no way to have nonzero \textbf{y}--directed electric components and \textbf{x}/\textbf{z}--directed magnetic components of the susceptibility tensors. This means that $\hat{\chi}_{yy}^{\textrm{ee}}$, $\hat{\chi}_{xx}^{\textrm{mm}}$, and $\hat{\chi}_{zz}^{\textrm{mm}}$ cannot be determined using TM-polarized probe waves,  and we consider them to be zero in the equations for the TM-polarized incidence. To find these components, one can use a TE-polarized wave as the incident field.
  \item Finally, the susceptibility components, which may arise in the case of the TM-wave incidence, are only three: $\hat{\chi}_{xx}^{\textrm{ee}}$, $\hat{\chi}_{zz}^{\textrm{ee}}$, and
  $ \hat{\chi}_{yy}^{\textrm{mm}}$. In general, it is not enough since we need also $\hat{\chi}_{yy}^{\textrm{ee}}$ and $ \hat{\chi}_{xx}^{\textrm{mm}}$. However, the particles have spherical symmetry and being arranged in a square array in the \textbf{xy}-plane, possess an in-plane isotropic response. As a result, we need to find only 3 components  $\hat{\chi}_{xx}^{\textrm{ee}} = \hat{\chi}_{yy}^{\textrm{ee}}$, $\hat{\chi}_{yy}^{\textrm{mm}} = \hat{\chi}_{xx}^{\textrm{mm}}$, and $\hat{\chi}_{zz}^{\textrm{mm}}$. The first two can be found from the study of the normal incidence. The third susceptibility $\hat{\chi}_{zz}^{\textrm{mm}}$ may be found applying an oblique TE incidence under an arbitrary angle.
\end{enumerate}

Considering the above hints, one may reduce the complex vectorial form of the equations \r{topbotBf:E}, \r{topbotBf:H} and \r{PMDyad:Mn} to a scalar problem. Therefore, the inverse problem of equation~\r{d4} may even be solved analytically as we did in Refs.~\onlinecite{Albooyeh1, Albooyeh6} for the general case when a metasurface is placed at an interface of two media with different permittivities. For a special case when the two surrounding media are the same, the results read as:
\begin{widetext}
\begin{eqnarray}
  \nonumber r &=& {{-{i \omega \over {2 \cos {\theta}}}[\eta_0 {\hat{\chi}_{xx}^{\textrm{ee}} \cos^2 {\theta} - \eta_0 \hat{\chi}_{zz}^{\textrm{ee}} \sin^2{\theta}- \hat{\chi}_{yy}^{\textrm{mm}}/\eta_0}]}\over{1-({\omega \over 2})^2 [{\hat{\chi}_{xx}^{\textrm{ee}}\hat{\chi}_{yy}^{\textrm{mm}}+ \eta_0^2 \hat{\chi}_{xx}^{\textrm{ee}} \hat{\chi}_{zz}^{\textrm{ee}} \sin^2 {\theta} }]-{{i \omega \over {2 \cos {\theta}}}[\eta_0 {\hat{\chi}_{xx}^{\textrm{ee}} \cos^2 {\theta} + \eta_0 \hat{\chi}_{zz}^{\textrm{ee}} \sin^2{\theta} + \hat{\chi}_{yy}^{\textrm{mm}}/\eta_0}]}}}, \l{RTth:R} \\
  \nonumber \\
  t &=& {{1+({\omega \over 2})^2 [{\hat{\chi}_{xx}^{\textrm{ee}}\hat{\chi}_{yy}^{\textrm{mm}}+ \eta_0^2 \hat{\chi}_{xx}^{\textrm{ee}} \hat{\chi}_{zz}^{\textrm{ee}} \sin^2 {\theta} }]}\over{1-({\omega \over 2})^2 [{\hat{\chi}_{xx}^{\textrm{ee}}\hat{\chi}_{yy}^{\textrm{mm}}+ \eta_0^2 \hat{\chi}_{xx}^{\textrm{ee}} \hat{\chi}_{zz}^{\textrm{ee}} \sin^2 {\theta} }]-{{i \omega \over {2 \cos {\theta}}}[\eta_0 {\hat{\chi}_{xx}^{\textrm{ee}} \cos^2 {\theta} + \eta_0 \hat{\chi}_{zz}^{\textrm{ee}} \sin^2{\theta} + \hat{\chi}_{yy}^{\textrm{mm}}/\eta_0}]}}}, \l{RTth}
\end{eqnarray}
\end{widetext}
and the retrieval formulas are as follows:

\begin{subequations}\l{SusFree}
     \begin{gather}
        \eta_0 \hat{\chi}_{xx}^{\textrm{ee}} = {-2i\over{\omega}}{{r_0+t_0-1}\over{r_0+t_0+1}}, \l{SusFree:EX}\\
        \nonumber \\
        {\hat{\chi}_{yy}^{\textrm{mm}} \over \eta_0} = {-2i\over{\omega}}{{r_0-t_0+1}\over{r_0-t_0-1}}, \l{SusFree:MY}\\
        \nonumber \\
        \sin^2{\theta} \eta_0 \hat{\chi}_{zz}^{\textrm{ee}} = {\hat{\chi}_{yy}^{\textrm{mm}}\over \eta_0} -{2i\over{\omega}}{{r_{\theta}-t_{\theta}+1}\over{r_{\theta}-t_{\theta}-1}}{\cos{\theta}}. \l{SusFree:EZ}
      \end{gather}
    \end{subequations}

Notice that in the above formulas the time dependence is assumed to be $\exp{(-i\omega t)}$. Also, $r_{0,\theta}$ and $t_{0,\theta}$ are, respectively, the reflection and transmission coefficients for $\theta_i=0, \theta$. The susceptibilities are retrieved from numerically simulated reflection and transmission data for two incidence angles $\theta_i=0, 10^\circ$. The susceptibility components are then plotted in Fig.~\ref{FIG_Plas_Sph}(b, c, d). As it is clear from the plots, the resonance amplitude of the magnetic susceptibility is two orders of magnitude smaller than those of the electric susceptibilities. It is in agreement with what we have expected, since there is no significant source for magnetic polarization in the metasurface. In order to examine whether these parameters can be used as characteristic parameters of the metasurface, the reflection amplitude and phase are plotted in Fig.~\ref{FIG_Plas_Sph}(e, f) for a different incident angle ($\theta_i=45^\circ$) using the retrieved susceptibilities from analytical formulas in \r{RTth} and \r{SusFree}. It is evident that the numerical results using HFSS simulation tool \cite{ref4_1_2} are very well matched with the predicted results from our retrieval. Moreover, the retrieved susceptibilities respect the constraints of causality and passivity \cite{ref4_1_2_0}; that is, the real parts of the susceptibility components are growing functions of frequency far from resonances and the sign of their imaginary parts does not change with the frequency and regarding to our assumption of the time dependence are always positive. The causal and passive behavior of the susceptibilities together with the correct prediction of reflection and transmission (not shown here) prove the applicability of the characterization model as well as its effectiveness.

So far, we successfully performed the characterization for a periodic metasurface. In the next section we show that the characterization model is not restricted to periodic metasurfaces. This model is applicable even for metasurfaces with disordered arrangements of  inclusions. In general, the applicability of our retrieval algorithm for disordered arrays keeps valid as long as the arrangement of particles is optically dense everywhere; that is, all inter-array particle distances in the random metasurface are sufficiently smaller than the operational wavelength.

\subsection{A planar array of coupled plasmonic nano-patches: a disordered metasurface}\label{DisMet}

\begin{figure}[h]
\centering
\includegraphics[width=0.49\textwidth,angle=0] {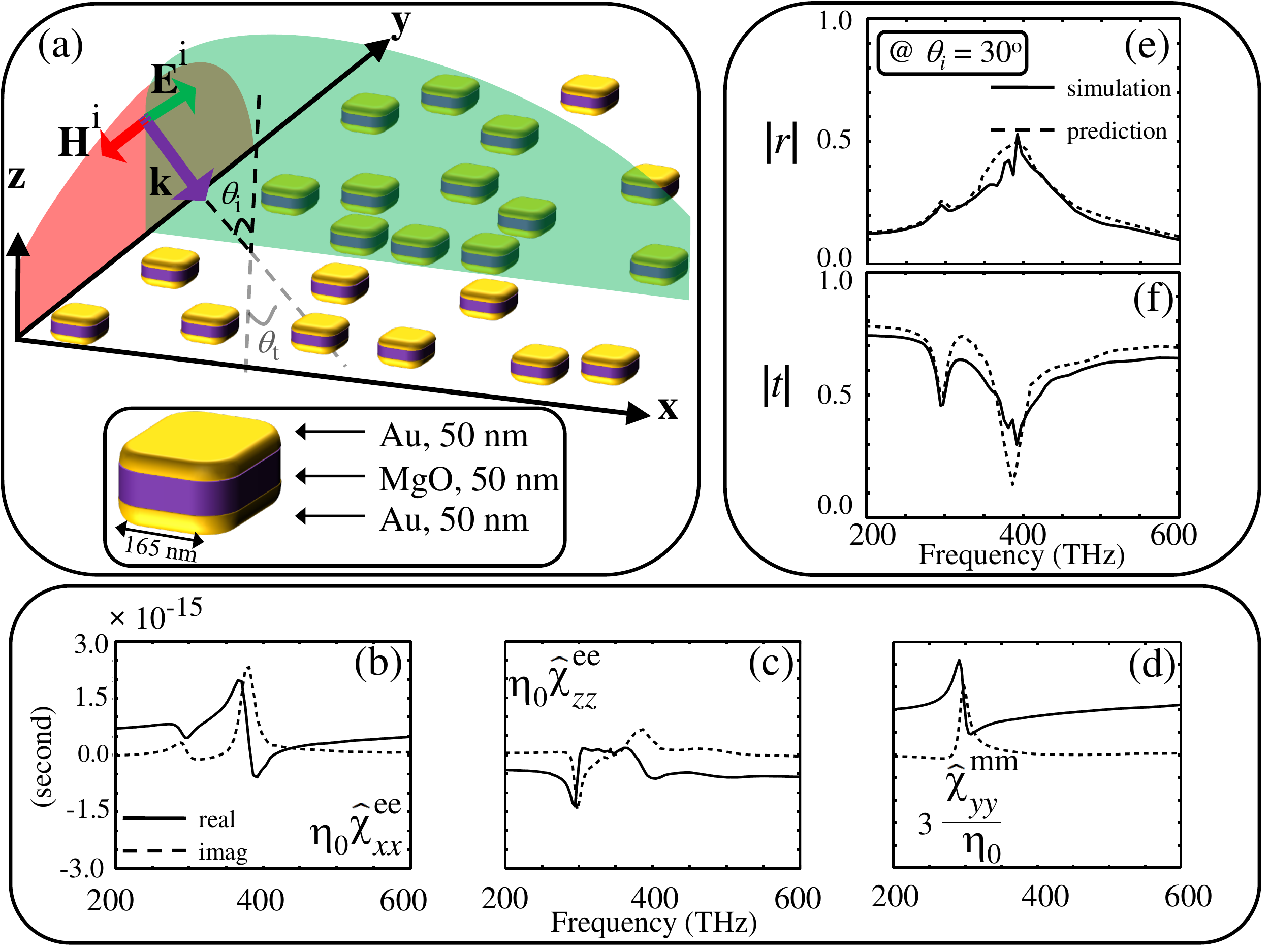}
\caption{(a) A planar disordered array of gold plasmonic coupled nano-patches located at the \textbf{xy}--plane on top of a fused silica substrate with the refractive index of $n=1.5$. The patches are separated by a dielectric spacer of MgO with the refractive index of $n=1.72$. The array is excited by a TM-polarized electromagnetic plane wave. (b) Tangential electric (c) Normal electric and (d) Tangential magnetic effective susceptibility components of the metasurface retrieved from the reflection and transmission data at $\theta_i=0$ and $45^\circ$ from numerical Finite Difference Time Domain method \cite{ref4_1_3, ref4_1_4}. (e) Predicted and simulated results for the amplitude of the reflection coefficient of the proposed metasurface for an incidence of $\theta_i=30^\circ$. (f) The same plot as in (e) for the amplitude of the transmission coefficient. The average unit cell size of the array is $510$~nm. The material data for gold is taken from \onlinecite{ref4_1}. Notice, the time dependence is assumed to be $\exp{(-i\omega t)}$ in these calculations.}
\label{FIG_Plas_Pat}
\end{figure}

In this section we present a different example of a metasurface: now it is composed of pairs of mutually coupled plasmonic nano-patches and the arrangement of these inclusions in the array is disordered. Figure \ref{FIG_Plas_Pat}(a) demonstrates the geometry of such an array. 
One may refer to work ~\onlinecite{ref4_2} to understand how the positional disorder is generated (when the disorder parameters varies from 0 to $\infty$ the array transits from a strictly regular to a fully  amorphous one). The effective susceptibilities are retrieved using the same approach as in the previous example and are plotted in Fig.\ref{FIG_Plas_Pat}(b, c, d).
This plot corresponds to an explicit value of the disorder parameter corresponding to a quite strong randomness (adequately illustrated by the drawing).
Due to the different type of inclusions, in contrast with the previous example, the magnetic response [Fig.\ref{FIG_Plas_Pat}(d)] is comparable with the electric response [Fig.\ref{FIG_Plas_Pat}(b, c)]. Indeed, it is shown in Ref.~\onlinecite{Albooyeh3} that the effective electric response is much more vulnerable to the positioning disorder compared to the magnetic response and the values of $\hat{\chi}_{xx}^{\textrm{ee}}$ are noticeably different from those which were retrieved for a periodic analogue of this metasurface.

Figures \ref{FIG_Plas_Pat}(e) and (f) show the prediction for the reflection and transmission coefficients for and incident angle of $\theta_i=30^\circ$ obtained through the retrieved characteristic parameters and directly by simulations. Indeed, the predicted results are in quite good agreement with the numerical results \cite{ref4_1_3, ref4_1_4} which demonstrates the predictive power of our approach in the characterization of disordered metasurfaces.

An important point is the anti-resonance behavior observed in the normal component of the electric susceptibility ($\hat{\chi}_{zz}^{\textrm{ee}}$) at the frequency range of the magnetic resonance ($\sim 300$~THz). This can be understood by recalling the averaged unit cell size of the array which is $510$~nm at this frequency; i.e., $a/\lambda \approx 0.5$. The anti-resonant feature is also slightly manifested in $\hat{\chi}_{xx}^{\textrm{ee}}$ at this frequency. To avoid the anti-resonance the high-order evanescent modes of the array must be included into the model which are out of the scope of our locally quasi-static study. More interesting is that at higher frequency ranges -- that of the electric resonance ($\sim 400$~THz), the array behaves as a homogeneous metasurface. That is, the passivity and causality constraints hold while the correct prediction of the reflection and transmission are also obtained. It is not very surprising, though the effective homogeneity is respected at even higher frequencies. The condition $a/\lambda \approx 0.5$  which holds at the magnetic resonance is, in fact, that of the spatial resonance of the array. At the corresponding frequencies the features of strong spatial dispersion are obviously enhanced. These features are non-Foster behavior of the real part and the wrong sign of the imaginary part of the retrieved material parameters. At the frequency of the electric resonance ($a/\lambda \approx 0.7$), the structure again becomes effectively homogeneous for the waves propagating under small angles to the normal. Beyond the lattice resonance and before condition of high-order propagating modes ($a/\lambda \ge 1$) holds, strong spatial dispersion can appear only because of a  noticeable phase shift per unit cell. However, we retrieved the characteristic parameters from data for $\theta_i \le 45^{\circ}$ and this shift is still sufficiently small.

With these two examples, we have shown the correctness and capability of the developed approach for the characterization of metasurfaces. In the next section, we characterize and discuss a bianisotropic metasurface with resonant magnetic response. We revisit an array of split-ring resonators (SRRs) as realizing a specific type of bianisotropy; i.e., omega-type bianisotropy. Moreover, we compare the characterization parameters retrieved from two different approaches; that is, HK and ST approaches. In the whole upcoming process, our characterization model shows off in the proof of the bianisotropy in the metasurface under study.


\subsection{Bianisotropic metasurfaces: SRRs make an omega-type bianisotropic metasurface}\label{BIMET}
As it is defined in the literature \cite{ref5_0_2} and was mentioned in section~\ref{chap2}, bianisotropy is understood as the electric response of a material to a magnetic excitation field and, vice versa, its magnetic response to an electric excitation field. This concept is formulated in equations \r{constRP} and \r{constRS} through  electromagnetic and magnetoelectric effective polarizability ($\hat{\bar{\bar{\alpha}}}^{\textrm{me}}$
and $\hat{\bar{\bar{\alpha}}}^{\textrm{em}}$) or susceptibility ($\hat{\bar{\bar{\chi}}}^{\textrm{me}}$
and $\hat{\bar{\bar{\chi}}}^{\textrm{em}}$) tensors as the characteristic parameters of bianisotropic metasurfaces, respectively. Bianisotropic metasurfaces can have many applications in various devices, including wave polarizers, absorbers, non-reciprocal surfaces, and planar antennas, to name only a few. Therefore, their analysis is very important and the milestone is to correctly characterize them in order to be able to design metasurfaces with desired optimal properties. According to Ref.~\onlinecite{ref5_0_2}, there are four general types of bianisotropy in materials; that is, reciprocal omega, reciprocal chiral, non-reciprocal Tellegen, and non-reciprocal ``moving''. There are also many approaches to design different type of bianisotropic elements and metasurfaces \cite{ref5_0_3, ref5_0_4}. In the present work we discuss one of these four classes, namely, reciprocal omega-type bianisotropy. In the tensor formalism, an omega-type bianisotropy has the following specifications \cite{ref5_0_2}:
\begin{itemize}
    \item If we decompose tensors $\hat{\bar{\bar{\chi}}}^{\textrm{me}}$ and $\hat{\bar{\bar{\chi}}}^{\textrm{em}}$ into symmetric and antisymmetric parts; i.e., $\hat{\bar{\bar{\chi}}}_{\textrm{sym}}= (\hat{\bar{\bar{\chi}}}+{\hat{\bar{\bar{\chi}}}}^\textrm{T})/2$ and $\hat{\bar{\bar{\chi}}}_{\textrm{asym}}= (\hat{\bar{\bar{\chi}}}-{\hat{\bar{\bar{\chi}}}}^\textrm{T})/2$, then the symmetric part is zero (because the metasurface is assumed to be non-chiral and exhibiting no pseudo-chirality effects).
  \item For reciprocal media, in general, we have: $\hat{\bar{\bar{\chi}}}^{\textrm{me}} = - {(\hat{\bar{\bar{\chi}}}^{\textrm{em}})}^\textrm{T}$.
\end{itemize}
In the above specifications, superscript ``$\textrm{T}$'' stands for the transpose of the corresponding tensor. Physically, for an omega-type bianisotropic metasurface, these specifications mean that the orientation of the magnetic moment which is induced by the electric field is orthogonal to the induced electric moment induced by the same electric field.

\begin{figure}[h]
\centering
\includegraphics[width=0.49\textwidth,angle=0] {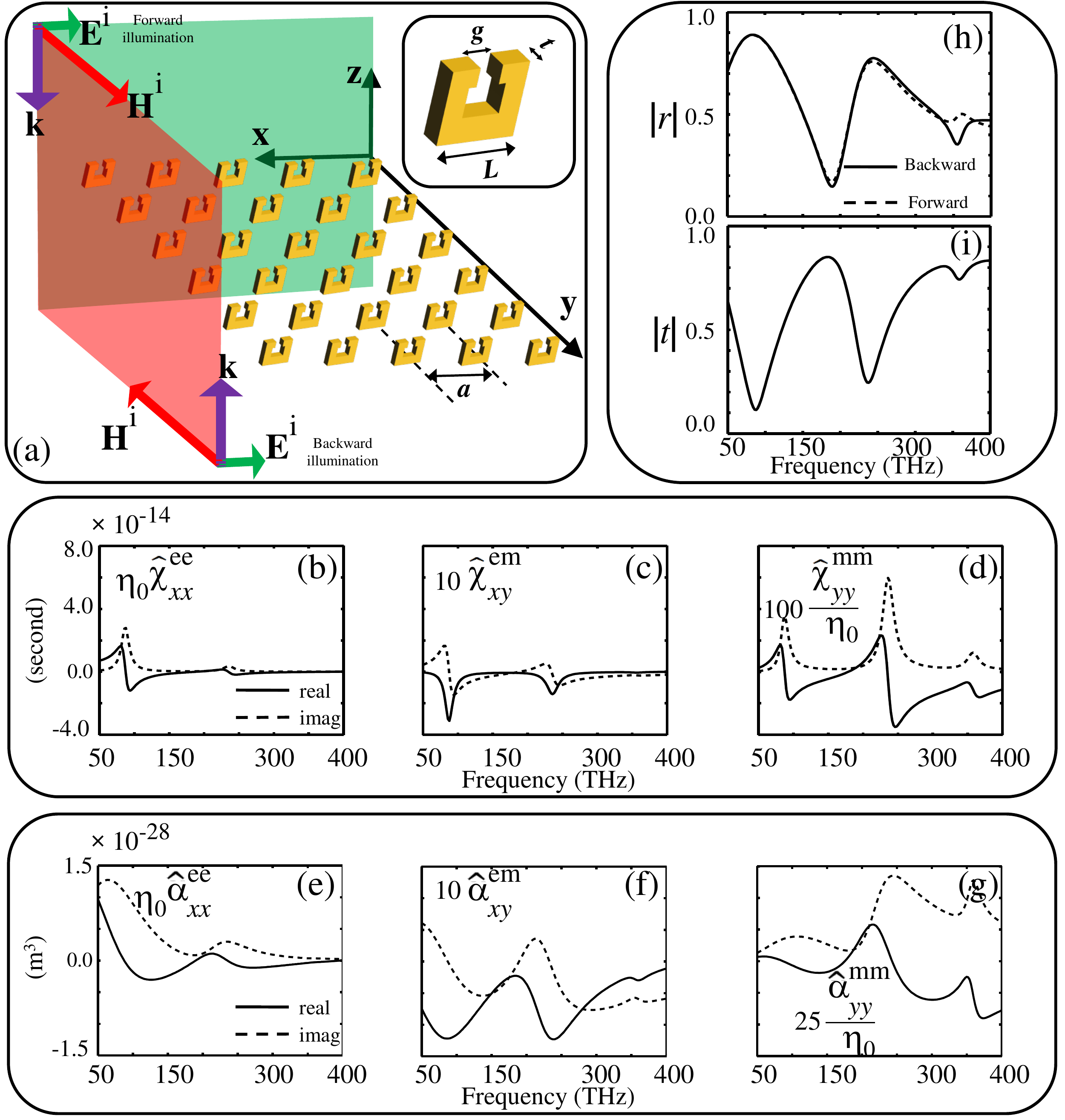}
\caption{(a) A planar periodic array of gold plasmonic split rings at the \textbf{xy}--plane located in free space. The array is excited by a normally-incident electromagnetic plane wave. (b) Tangential electric (c) Magneto-electric and (d) Magnetic effective susceptibility components of the metasurface retrieved from the reflection and transmission data using the HK approach. (e), (f), and (g) Similar curves for collective polarizabilities calculated using the ST model. (h) Simulated results for the amplitude of the reflection coefficients of this metasurface for forward and backward directions. (i) Simulated result for the amplitude of the transmission coefficient. The unit-cell size of the array is $a=175$~nm. The other geometrical dimensions of each inclusion are: $g=80$~nm, $t=30$~nm, and $L=170$~nm. The material data for gold is taken from \onlinecite{ref5_0_1}. Notice the factors 10, 25 and 100 in front of the effective magnetoelectric and magnetic susceptibilities/polarizabilities, respectively. The time dependence is $\exp{(-i\omega t)}$.}
\label{FIG_Plas_SRR}
\end{figure}

\textbf{A metasurface of SRRs}: Let us consider a metasurface composed of split-ring resonators [Fig.~\ref{FIG_Plas_SRR}(a)].
The split-ring resonators shown in this figure exhibit an omega-type bianisotropy, since the magnetic polarization induced by applied electric field is orthogonal to the direction of the applied field (the same holds for the electric polarization induced by applied magnetic fields). We excite such metasurface by an electromagnetic wave with its magnetic field vector $\-H^\textrm{i}$ along the ring axis. This excitation creates a circulating current which in turn produces a net magnetic moment that causes a magnetic resonant response. To reduce the complexity of the problem and with the aim to retrieve tangential effective susceptibilities or polarizabilities, we excite this array by a normally incident plane wave. The goal in this example is to show the differences between two characterization approaches of HK and ST in the characterization of metasurfaces. We may consider all $36$ susceptibility/polarizability parameters to fully  characterize our metasurface of omega-type array of split-ring resonators. However, this would be very complicated and make the problem difficult to solve. Therefore, we make use of all our a priori knowledge in making the problem as simple as possible with the smallest number of unknown characteristic parameters. According to the discussion in Section \ref{periMS}, the only nonzero susceptibility terms (the HK model) associated with the specific excitation shown in Fig.~\ref{FIG_Plas_SRR}(a) would be $\hat{{{\chi}}}^{\textrm{ee}}_{\textrm{xx}}$, $\hat{{{\chi}}}^{\textrm{mm}}_{\textrm{yy}}$, and $\hat{{{\chi}}}^{\textrm{em}}_{\textrm{xy}} = -\hat{{{\chi}}}^{\textrm{me}}_{\textrm{yx}}$ (or, equivalently, the nonzero collective polarizabilities of the ST model are $\hat{{{\alpha}}}^{\textrm{ee}}_{\textrm{xx}}$, $\hat{{{\alpha}}}^{\textrm{mm}}_{\textrm{yy}}$, and $\hat{{{\alpha}}}^{\textrm{em}}_{\textrm{xy}} = -\hat{{{\alpha}}}^{\textrm{me}}_{\textrm{yx}}$). Using our homogenization model, we may solve the problem for the normal incidence to find the reflection and transmission coefficients as:
\begin{eqnarray}
  \nonumber r_{\pm} &=& {-{i \omega \over 2}[\eta_0 {\hat{\alpha}_{xx}^{\textrm{ee}} \pm (\hat{\alpha}_{yx}^{\textrm{me}}-\hat{\alpha}_{xy}^{\textrm{em}}) - \hat{\alpha}_{yy}^{\textrm{mm}}/\eta_0}]}, \l{RTBi:R} \\
  \nonumber \\
   t &=& 1{-{i \omega \over 2}[\eta_0 {\hat{\alpha}_{xx}^{\textrm{ee}} + \hat{\alpha}_{yy}^{\textrm{mm}}/\eta_0}]}, \l{RTBiST}
\end{eqnarray}
for ST model and
\scriptsize
\begin{eqnarray}
  \nonumber r_{\pm} &=& {{-{i \omega \over 2}[\eta_0 {\hat{\chi}_{xx}^{\textrm{ee}} \pm (\hat{\chi}_{yx}^{\textrm{me}}-\hat{\chi}_{xy}^{\textrm{em}}) - \hat{\chi}_{yy}^{\textrm{mm}}/\eta_0}]}\over{1-({\omega \over 2})^2 [{\hat{\chi}_{xx}^{\textrm{ee}}\hat{\chi}_{yy}^{\textrm{mm}}- \hat{\chi}_{yx}^{\textrm{me}} \hat{\chi}_{xy}^{\textrm{em}} }]-{i \omega \over 2}[{\eta_0 \hat{\chi}_{xx}^{\textrm{ee}}+\hat{\chi}_{yy}^{\textrm{mm}}/\eta_0}]}}, \l{RTBi:R} \\
  \nonumber \\
  t &=& {{1+({\omega \over 2})^2 [{\hat{\chi}_{xx}^{\textrm{ee}}\hat{\chi}_{yy}^{\textrm{mm}}- \hat{\chi}_{yx}^{\textrm{me}} \hat{\chi}_{xy}^{\textrm{em}} }]}\over{1-({\omega \over 2})^2 [{\hat{\chi}_{xx}^{\textrm{ee}}\hat{\chi}_{yy}^{\textrm{mm}}- \hat{\chi}_{yx}^{\textrm{me}} \hat{\chi}_{xy}^{\textrm{em}} }]-{i \omega \over 2}[{\eta_0 \hat{\chi}_{xx}^{\textrm{ee}}+\hat{\chi}_{yy}^{\textrm{mm}}/\eta_0}]}}, \l{RTBiHK}
\end{eqnarray}
\normalsize
for the HK approach. Note that $r_+$ and $r_-$ stand for the reflection from the forward and backward directions, respectively, and $t$ is the transmission coefficient. Also, for the reciprocal omega-type bianisotropy $\hat{\chi}_{yx}^{\textrm{me}} = - \hat{\chi}_{xy}^{\textrm{em}}$. Moreover, due to the geometrical asymmetry, the reflection coefficients are different when the metasurface is illuminated from the forward and backward directions. If we carelessly do not consider this asymmetry, then we may neglect an important factor in the correct characterization. For example, if we use the same characterization model as we did in the first example of the previous section (identical coupled nano-patches), then we may obtain nonphysical characteristic parameters. Moreover, even if we get physical retrieved parameters from the forward data, they would not correctly predict the reflection of the metasurface for the backward direction \cite{ref5_0_5}. Therefore, we must add the bianisotropic terms $\hat{{{\chi}}}^{\textrm{em}}_{\textrm{xy}}$ and $\hat{{{\chi}}}^{\textrm{me}}_{\textrm{yx}}$ (or equivalently $\hat{{{\alpha}}}^{\textrm{em}}_{\textrm{xy}}$ and $\hat{{{\alpha}}}^{\textrm{me}}_{\textrm{yx}}$) in our characterization model in order to achieve physical results. Solving equations \r{RTBiST} and \r{RTBiHK} for the collective polarizability or effective susceptibility components leads to:
\begin{eqnarray}
  \nonumber \eta_0 \hat{\alpha}_{xx}^{\textrm{ee}} &=& -{i\over{\omega}}\left[{{1-t-{(r_++r_-)\over 2}}}\right],
\l{SusBi:EXST} \\
  \nonumber {\hat{\alpha}_{yy}^{\textrm{mm}}\over \eta_0 } &=& -{i\over{\omega}}\left[{1-t+{(r_++r_-)\over 2}}\right],
\l{SusBi:MYST} \\
\hat{\alpha}_{xy}^{\textrm{em}} &=& - \hat{\alpha}_{yx}^{\textrm{me}} = -{i\over{\omega}} \left({{r_+ - r_-}\over 2}\right),
\l{SusBi:MEST}
\end{eqnarray}
for the collective polarizabilities of the ST model and
\begin{eqnarray}
  \nonumber \eta_0 \hat{\chi}_{xx}^{\textrm{ee}} &=& -{2i\over{\omega}}\left[{1-{\Delta} \left({1+t-{{r_++r_-}\over 2}}\right)}\right],
\l{SusBi:EXHK} \\
  \nonumber {\hat{\chi}_{yy}^{\textrm{mm}}\over \eta_0 } &=& -{2i\over{\omega}}\left[{1-{\Delta} \left({1+t+{{r_++r_-}\over 2}}\right)}\right],
\l{SusBi:MYHK} \\
\hat{\chi}_{xy}^{\textrm{em}} &=& - \hat{\chi}_{yx}^{\textrm{me}} = -{2i\over{\omega}} \Delta \left({{r_+ - r_-}\over 2}\right),
\l{SusBi:MEHK}
\end{eqnarray}
for the effective susceptibilities of the HK model. In equations \r{SusBi:MEHK}, parameter $\Delta$ denotes
\e
        \Delta = {2 \over {(1+t)^2+\left({{r_+ - r_-}\over 2}\right)^2-\left({{r_+ + r_-}\over 2}\right)^2}}.\l{SusBi:Delta}
        \f
As it is clear, the extraction of the susceptibilities of the HK model is more complicated than the extraction of collective polarizabilities of the ST approach. The retrieved susceptibility/polarizability components for an array of split-ring resonators are given in Fig.~\ref{FIG_Plas_SRR}(b, c, d, e, f, g). Note that the HK model results in more distinguishable resonances compared with the ST approach. However, both results sound physical since they obey the causality and passivity constraints. The presence of bianisotropic response is obvious from Fig.~\ref{FIG_Plas_SRR}(c, f).


Indeed, the magnetic polarization is obviously resonant and mainly resulting from the magnetoelectric susceptibility/polarizability which is by one order of magnitude higher than the magnetic one (normalized to $\eta_0$) [see Fig.~\ref{FIG_Plas_SRR}(c, d) or (f, g)]. Again, one may not explicitly claim a magnetoelectric and/or magnetic coupling from only the reflection or transmission data in Fig.~\ref{FIG_Plas_SRR}(h) or (i). The magnetic response and magnetoelectric coupling are not distinguishable a priori, in wave-excitation scenarios. To understand which of them is dominating one needs to correctly characterize the metasurface. And this tool is given by the presented characterization method. It turns out that the magnetic susceptibility/polarizability of the array of SRRs is pretty small. This result fits the known theoretical models which show that the magnetic response of metal split rings is a second-order spatial dispersion effect while the magnetoelectric coupling effect is a (generally) stronger, first-order effect in  terms of the electrical size of the particle.
Another issue regards the asymmetric reflection for the illumination from forward and backward directions [Fig.~\ref{FIG_Plas_SRR}(h)]. In the present case, when the substrate is absent, this asymmetry is related to the bianisotropy in the metasurface inclusions.


\section{Conclusion}
Starting from a general overview of the homogenization and characterization of metasurfaces, we have presented a brief history review of this topic and have identified two important and widely used approaches in this field. We have then analytically generalized these two approaches for a general bianisotropic metasurface supported by a refractive substrate. We have started the theory by the derivation of sheet transition conditions. Thereafter, we have continued to find the reflection and transmission dyadics in terms of the surface electric and magnetic dipole polarizations. Finally, we have presented the reflection and transmission dyadics in terms of the effective susceptibilities for the HK model and in terms of the collective polarizabilities for the ST approach. We have next demonstrated the use of the homogenization theory with a periodic as well as an amorphous metasurface. Finally, we have presented an example of a bianisotropic metasurface and compared the two approaches together.
We believe that the developed general homogenization framework can be successfully used in modeling and design of metasurfaces for various applications in different frequency ranges. Moreover, it will help to gain physical insight into the physical properties of complex metasurfaces.

\section*{Acknowledgements}
We would like to thank Dr. C. Menzel and Mr. M. Yazdi who provided us with  some of the numerical simulations for the amorphous and for the split-ring resonator arrays, respectively.

\end{document}